\documentclass[twocolumn,showpacs,preprintnumbers,amsmath,amssymb]{revtex4}
\usepackage{graphicx,psfrag}
\usepackage{dcolumn}
\usepackage{bm}

\begin{document}

\preprint{Manuscript}

\title{Predictive protocol of flocks with small-world connection pattern}

\author{Hai-Tao Zhang$^1$}
\author{Michael Z. Q. Chen$^{1,2,\dagger}$}
\author{Tao Zhou$^{3,4}$}
\affiliation{$^1$Department of Engineering, University of Cambridge, Cambridge CB2 1PZ, U.K. \\
$^2$Department of Engineering, University of Leicester, Leicester
LE1 7RH,
U.K. \\
$^3$Department of Modern Physics, University of Science and Technology of China, Hefei 230026, PR China\\
$^4$Department of Physics, University of Fribourg, Chemin du Muse 3, Fribourg CH-1700, Switzerland}


\begin{abstract}
By introducing a predictive mechanism with small-world connections,
we propose a new motion protocol for self-driven flocks. The
small-world connections are implemented by randomly adding
long-range interactions from the leader to a few distant agents,
namely pseudo-leaders. The leader can directly affect the
pseudo-leaders, thereby influencing all the other agents through
them efficiently. Moreover, these pseudo-leaders are able to predict
the leader's motion several steps ahead and use this information in
decision making towards coherent flocking with more stable
formation. It is shown that drastic improvement can be achieved in
terms of both the consensus performance and the communication cost.
From the industrial engineering point of view, the current protocol
allows for a significant improvement in the cohesion and rigidity of
the formation at a fairly low cost of adding a few long-range links
embedded with predictive capabilities. Significantly, this work
uncovers an important feature of flocks that predictive capability
and long-range links can compensate for the insufficiency of each
other. These conclusions are valid for both the attractive/repulsive swarm model and the Vicsek model.\end{abstract}


\maketitle

\section{Introduction}
Over the last decade, physicists have been looking for common,
possibly universal, features of the collective behaviors of
animals, bacteria, cells, molecular motors, as well as driven
granular objects. The collective motion of a group of autonomous
agents (or particles) is currently a subject of intensive research
that has potential applications in biology, physics and
engineering. One of the most remarkable characteristics of systems,
such as flocks of birds, schools of fish, and swarms of locusts, is
the emergence of collective states in which the agents move in the
same direction, i.e., an ordered state
\cite{vi95,do06,ch06,he05,gr04,al07,li08}. Moreover, this ordered state
seeking problem for flocks/swarms/schools can be further
generalized to consensus \cite{sa04}, rendezvous, synchrony,
cooperation and so on. From the application aspect, this kind of
distributed collective dynamic systems has direct implications on
sensor network data fusion, load balancing, unmanned air vehicles
(UAVs), attitude alignment of satellite clusters, congestion
control of communication networks, multi-agent formation control
and global coordination for emergency \cite{ak02,og04,ar02,he00}.

The interaction pattern of the natural biological flocks/swarms are
neither entirely regular nor entirely random. An individual of a
flock usually knows its neighbors, but its circle of acquaintances
may not be confined to those who live right next door. In 1998, in
order to describe the transition from a regular lattice to a random
graph, Watts and Strogatz (WS) introduced the concept of the
small-world network \cite{wa98} by rewiring one end of a few
connections to new nodes chosen at random from the whole network.
With these few shortcuts, the average distance is decreased
significantly without crucially changing the clustering property.
The work on the WS small-world network has started an avalanche of
research on complex networks, especially, the synchronizability of
networks can be greatly enhanced by introducing a few long-range
connections \cite{wa03,zh06,al03}. Thus, for better
synchronization in a flock of neighboring-connected agents with a leader, it is advantageous to build a small-world-type
network structure by randomly adding long-range connections from
the leader to a few distant agents (namely \textit{pseudo-leaders}),
so that the leader can affect them, thereby influencing all the
other agents through them, via fast communication and rapid control
commands.

Although a lot of relevant works were focused on network structures,
recently, more and more researchers are interested in finding the
rules of the inter-connections present in abundant bio-groups.
Extraction of these rules can help interpret why the bio-groups can
demonstrate so many good characteristics such as synchronization,
stabilisation, cohesion, etc. A fairly basic but popular flocking
strategy can be traced back to the Reynolds Model \cite{re87}, in
which three elementary flocking protocols are prescribed, (i)
\textit{separation}: steer to avoid crowding and {collision;} (ii)
\textit{alignment}: steer towards the average {heading;} (iii)
\textit{cohesion}: steer to move towards the average position. These
rules have been proven to be effective and thus become the basic
rules for the design of bio-group dynamic models. In 2003, Gazi and
Passino \cite{ga03} proposed an effective A/R (attractive/repulsive)
swarm model in which the motion of each individual (autonomous agent
or biological creature) is determined by two factors: (i) attraction
to the other individuals on long distances; (ii) repulsion from the
other individuals on short distances.

The Gazi and Passino A/R model \cite{ga03}, embedded with a similar
mechanism of the inter-molecular force, is derived from the
biological flocks/swarms behaviors. Thus far, the general understanding is
that the swarming behaviors result from an interplay between a
long-range attraction and a short-range repulsion between the
individuals \cite{br54,wa91}. In \cite{br54}, Breder suggested a
simple model composed of a constant attraction term and a repulsion
term which is inversely proportional to the square of the distance
between two members, whereas in \cite{wa91} Warburton and Lazarus
studied the effects on cohesion of a family of attraction/repulsion
functions. Moreover, in physics communiuty, a large volume of
literature on systems with interactive particles have also adopted functions of
attractive and repulsive forces to investigate the dynamics
of the system \cite{le98,be90,we98,ar07,sc92,me99,oh93}. For
instance, in \cite{le98,be90}, systems of particles interacting in a
lattice are considered with attraction between particles located at
different sites and repulsion between particles occupying the same
site. It is discussed in \cite{we98} that, the structure of a
nonuniform Lennard-Jones (LJ) liquid near a hard wall is
approximated by that of a reference fluid with repulsive
intermolecular forces in a self-consistently determined external
mean field incorporating the effects of attractive forces.

With the A/R model \cite{ga03}, Gazi and Passino proved that the
individuals will form a bounded cohesive swarm in a finite time. One
year later, by adding another factor, i.e., attraction to the more
favorable regions (or repulsion from the unfavorable regions), they
generalized their former model into a social foraging swarm model
\cite{ga04}. Under some suitable circumstance, agents in this
modified model are apt to move to the more favorable regions. The A/R model has been
adopted by physicists and biologists to model self-driven particles
and biological flocks \cite{sh06,zi07,do06,ch07,li07a,li07b}.  In
2004, Moreau \cite{mo04} presented a linearized model of flocks, and
proved that the flock is uniformly and globally cohesive to a
bounded circle if and only if there exists an agent (the ``leader")
connecting to all other agents, directly or indirectly, over an
arbitrary time interval. Other than these kinds of A/R models
\cite{ga03,ga04,mo04}, another popular kind of model is that without leaders
(say homogeneous), where a very representative one is the Vicsek
Model \cite{vi95}. In each step, every agent updates its velocity
according to the average direction of its
neighbors. With the decrease of external noise
or the increase of the particle density, the collective behavior of
the flock undergoes a phase transition from the randomly distributed
phase to the coherently moving phase. In 2003, Jadbabaie {\it et al.}
provided the convergence condition of the noise-free Vicsek model,
i.e., the individuals are linked in some intervals \cite{ja03}.

\begin{figure}
\scalebox{0.40}[0.40]{\includegraphics{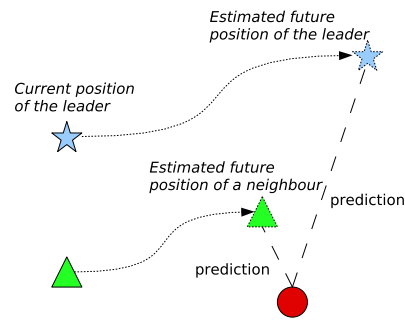}} \caption{ (Color
online) Predictive vision in natural bio-groups. Each individual
makes its motion decision based on not only the current status of
the leader and its neighbors but also their future dynamics. }
\label{fig: Predictive vision}
\end{figure}

Although most of the previous works on flock dynamics
yield many advantages such as synchronization, stabilization,
cohesion, and quick consensus, agents within the networks only know
the information that is currently available to them. In this paper,
we highlight another appealing phenomenon, i.e. the universal
existence of predictive mechanism in various biological aggregated
systems. A general physical picture behind this is illustrated in
Fig.~\ref{fig: Predictive vision} and interpreted as follows: in
widely-spread natural bio-groups composed of animals, bacteria,
cells, etc., the decision  on the next-step behavior of each
individual is not solely based on the currently available state
information (including position, velocity, etc.) of other
(neighboring) agents inside the group but also on the predictions of
future states. More precisely, taking a few past states of its
leader and neighbors into account, an individual can estimate the
corresponding future states several steps ahead and then make a
decision. 

Some experimental evidences have already been reported in the
literature. In 1959, Woods implemented some experiments on bee
swarms and found a certain predictive mechanism of electronic
signals inside this bio-group \cite{wo59}. Also for bee swarms, in
2002, Montague {\it et al.} discovered that there exist some
predictive protocols in the foraging process in uncertain
environments \cite{mo00}. Apart from the investigation of the
predictive mechanisms of locust swarming and foraging, more
scholars focused on the predictive function of the optical and
acoustical apparatuses of the individuals inside bio-groups
\cite{go03,su06,me07}, especially cortexes and retinae. For
instance, based on intensive experiments on the bio-eyesight
systems, they found that when an individual observer prepared to
follow a displacement of the stimulus with the eyes, visual form
adaptation was transferred from current fixation to the future
gaze position. These investigations strongly support our
conjecture of the existence of some predictive mechanisms inside
abundant bio-groups.

Bearing in mind the plentiful examples of predictive protocols
inside natural bio-groups, we incorporated some predictive functions
into a few long-range links, and found that it is possible to
significantly enhance the flocking performances at a fairly low
cost of the additional predictive energy. More interestingly, proper
prediction capability can help reduce the minimal number of the
long-range links between the leader and the pseudo-leaders, thus
effectively decrease the communication cost.

On the other hand, from the industrial application point of view,
the phenomena and strategy reported in this paper may be applicable
in some relevant prevailing engineering areas like autonomous robot
formations, sensor networks and UAVs \cite{ak02,og04,ar02}.
Typically, due to the limitation of the communication energy, only a
few agents have the capability to communicate with the leader. The
incorporation of a predictive mechanism into these pseudo-leaders
can greatly improve the flocking performances.

The rest of this paper is organized as follows. In
Section~\ref{sec: model}, the small-world
connection model with embedded predictive mechanism is presented. Then, in Section~\ref{Sec: AR results}, its important role in improving flocking synchronization
performances is extracted and analyzed by numerical simulations on the
A/R model \cite{ga03}. Afterwards, in Section~\ref{Sec: prediction
Vicsek}, the generality of the virtues endowed by such predictive
mechanism is validated on the Vicsek model \cite{vi95} as well. Finally,
conclusions are drawn in Section~\ref{sec: conclusion}.

\section{Model}\label{sec: model}
\begin{figure}[htb]
\centering \leavevmode {\includegraphics[width=7.0cm]{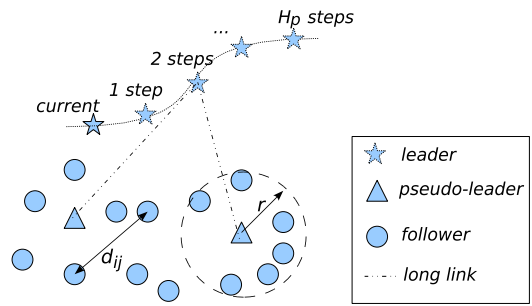}}
\caption{Small-world predictive mechanism of flocks. Here, the
leader (L), pseudo-leaders (P), and followers (F) are denoted by
star, triangles and circles, respectively, and the neighboring area
of each individual is a circle with radius $r$ centering at itself.
The leader's trajectory is given in advance, which will not
influenced by the others. Each pseudo-leader's dynamics is always
influenced by both the leader's future position $H_p$ steps ahead and
its neighbors' positions, while each follower's dynamics is solely
affected by its neighbors' positions.} \label{fig:structure}
\end{figure}

It is well-known that a ring-shaped network structure is not a good
one for efficient mutual communication and  global control within a
flock of agents, while the so-called small-world networking
structure performs much better. By adding a few
long-range connections, the average path length of the ring-shaped
network will be abruptly decreased. This small-world effect is very
desirable for fast communication and information transmission,
efficient synchronization, and effective global control over the
entire network \cite{zh06}. Thus, in a flock of
neighboring-connected agents with a leader, for communication and
control purposes, it is advantageous to build a small-world-type
network by randomly adding long-range connections from the leader to
a few distant agents (namely \emph{pseudo-leaders}). As shown in
Fig.~\ref{fig:structure}, the leader can affect pseudo-leaders,
thereby influencing all the other agents through them. To be clear,
we call the non-special agents \emph{followers}. Thus in our model,
there are three different kinds of agents: leader (L),
pseudo-leaders (P), and followers (F).

\begin{figure}[htb]
\centering \leavevmode {\includegraphics[width=6.0cm]{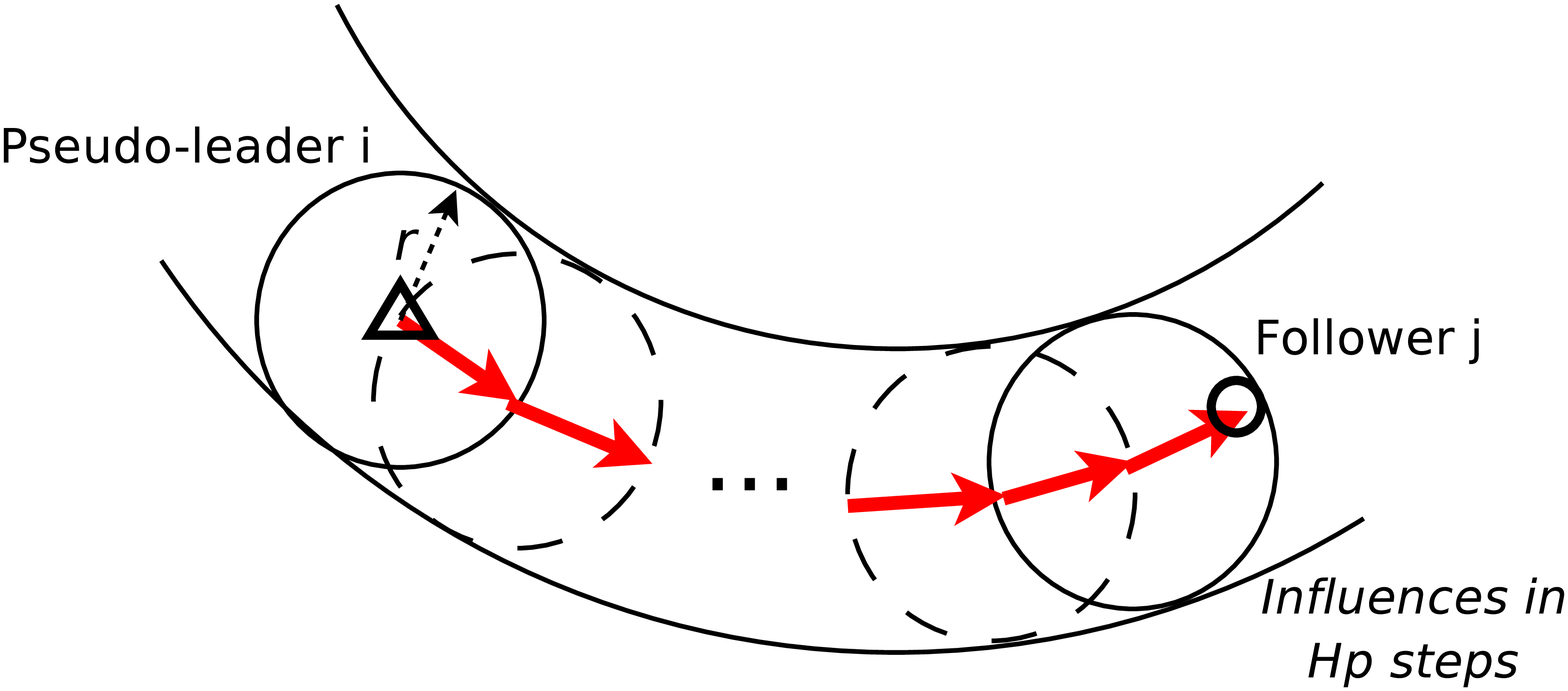}}
\caption{(Color online) Information communication process inside
flocks with a predictive mechanism. The arrows represent the passing
of position information. If the pseudo-leader $i$ predicts the
dynamics of the leader $H_p$ steps ahead, then the follower $j$,
within topological distance equaling $H_p$ from $i$, could be
affected by the current location of the leader. Here, an agent $j$
has topological distance zero to itself, and 1 to all the agents
located inside the circle with radius $r$ centered on $j$. Two
agents having topological distance 1 are called connected. An agent
has topological distance $D_T$ to $j$ if and only if it is connected
to at least one agent with distance $D_T-1$ to $j$, while is not
connected with any agents having distance smaller than $D_T-1$ to
$j$. Note that all the agents are not necessarily embedded in a
circular pathway. Instead, due to the circle-shaped neighborhood of
each agent, the position information transmission pathway forms a
tube tangent to these neighboring circles, and the layout boundary
of the pathway could be any types of curves depending on the
location of the agents. } \label{fig:influence}
\end{figure}

In this model, the flock is assumed to move in an $m$-dimensional
space and the standard A/R function \cite{sa04,ga03,ga04}
\begin{equation}
\label{eq: AR function} G(d_{pL})=-d_{pL}\left(a-b\cdot
\mbox{exp}(-\|d_{pL}\|_2^2/c)\right)
\end{equation}
is used as long-range interaction from the leader (L) to each
pseudo-leader (P), where $a$, $b$, $c$ are three free parameters,
$d_{pL}$ is the $m$-dimensional vector pointing from the predicted
location of leader $L$ to the current location of a pseudo-leader
$p$, and $\|d_{pL}\|_2=\sqrt{d_{pL}^Td_{pL}}$ denotes the Euclidean
distance between them. The force $G(d_{pL})$ is an $m$-dimensional
vector whose direction is from the pseudo-leader to the leader. For
simplicity, in our model, the motion of leader is given in advance,
and will not be affected by any other agents. We assume every
pseudo-leader has the same prediction horizon $H_p$, that is to say,
a pseudo-leader will predict the leader's location $H_p$ steps
ahead.

On the other hand, a weaker A/R function, representing the
short-range interaction between two arbitrary neighboring agents
$i$ and $j$ is addressed as:
\begin{equation}
\label{eq: AR function of followers}
g(d_{ij})=-d_{ij}\left(\tilde{a}-\tilde{b}\cdot
\mbox{exp}(-\|d_{ij}\|_2^2/\tilde{c})\right),
\end{equation}
where $d_{ij}$ is the $m$-dimensional vector pointing from the
individuals $j$ to $i$, and $\|d_{ij}\|_2=\sqrt{d_{ij}^Td_{ij}}$
denotes the Euclidean distance between them. The parameters
$\tilde{a}$, $\tilde{b}$ and $\tilde{c}$ are much smaller than $a$,
$b$, and $c$, respectively. The direction of vector $g(d_{ij})$ is
from $i$ to $j$. Denote by $r$ the radius of neighboring area (see
Fig.~\ref{fig:structure}). The neighboring A/R links could connect
any two agents (F-F, P-P and L-F) within the Euclidean distance $r$
except the L-P interaction described in Eq.~({\ref{eq: AR
function}}). Note that the leader can influence other agents, but
will not be influenced. In order to decrease the prediction cost, no
predictive mechanism is incorporated into the neighboring A/R links.
Bearing in mind the physical meaning of A/R function \cite{sa04},
the positions of a pseudo-leader $z_p$ and a follower $z_i$ (both
$z_p$ and $z_i$ are $m$-dimensional vectors) are determined by
\begin{equation}
\label{eq: pseudo-leader dynamics}
\dot{z}_{p}(t)=\underbrace{G(d_{pL}(t+H_p))}_{long~link~to~the~
leader}+\underbrace{\sum_{j\neq L, d_{pj}(t)\leq
r}g(d_{pj}(t)),}_{neighboring~links}
\end{equation}
and
\begin{equation}
\label{eq: follower dynamics}
\dot{z}_i(t)=\underbrace{\sum_{j,d_{ij}(t)\leq
r}g(d_{ij}(t)),}_{neighboring~links}
\end{equation}
respectively, where $t$ denotes the current time, and
$d_{pL}(t+H_p)$ represents the $m$-dimensional vector pointing from
the leader's position $H_p$ steps ahead to the current position of a
pseudo-leader. Although the rest of this paper concentrates on the motions in
$2$-dimensional space, the present model can be directly applied in
any finite dimensional space. In this way, unlike the routine
flocking strategies \cite{mo04,ga03,ga04,co05}, a small-world
interaction pattern is established with embedded predictive
mechanism, which has the capability of predicting the future
behavior (position, velocity, etc.) of the leader several steps
ahead. Note that the structure of this interaction network will
change in time since the location of each agent is varying.

The information communication process is illustrated in
Fig.~\ref{fig:influence}. The farthest agent $i_1$ directly
communicating with agent $i$ is among the ones at the rim of the
circle with radius $r$ centered on agent $i$. Analogously, the
farthest agent $i_2$ directly influenced by $i_1$ is also located at
the rim of the circle centered on agent $i_1$, and so forth.
Finally, the influence of agent $i$ reaches agent $j$ in $H_p$
steps. When agent $j$ receives the information from agent $i$ at
time $t$, it is in fact a delayed information of agent $i$ at time
step $t-H_p$. However, if agent $i$ acts as a pseudo-leader who can
accurately predict the behavior of the leader $H_p$ steps ahead,
then, at time step $t$, agent $j$'s motion is affected by the exact
current location of the leader $z_L(t)$. In this way, although agent
$j$ may not have direct connection with the leader, it could know
some information of the leader's current dynamics by agent $i$'s
delayed information. Therefore, agent $j$ can adhere to the leader
more tightly, the flock's formation is more likely to be stable, and the coherence of the whole flock is thus improved
effectively. Note that, the predictive mechanism is valid only if
the leader's motion is regular \cite{ex1}. If the leader moves in
some random, chaotic or  other irregular ways, such as
random walk, it is, in principle, impossible for the pseudo-leaders
to predict the leader's further location. Fortunately, in
the real biological world, the flock leader always moves in some
predictable pattern. Therefore, the other agents have the opportunity
to display their predicting ability, such as that used by a chameleon
to capture a fly and by a dog to catch a frisbee.

\section{Analysis and simulations}\label{Sec: AR results}
To show the advantages of the predictive mechanism, we compare the
performances of the two cases of flocking with and without the
predictive mechanism by simulations over an $N$-agent flock moving
in a two-dimensional space as shown in Fig.~\ref{fig:final position
with predictive mechanism}. The parameters are set as follows: the
neighboring circle $r=0.65$, the parameters of the A/R functions
(see Eqs.~(\ref{eq: AR function}) and (\ref{eq: AR function of
followers})) of long-range links and neighboring links are set as
$a=8$, $b=17.6$, $c=3.2$, and $\tilde{a}=1$, $\tilde{b}=2.2$,
$\tilde{c}=0.2$, respectively. The former A/R function is much
stronger in order to intensify the influence of the leader. As shown
in Fig.~\ref{fig:final position with predictive mechanism}a, each
agent starts from a position randomly selected in the square
$[0,1]\times [0,1]$. The leader and the pseudo-leaders are selected
randomly among these $N$ agents. The
trajectory of the leader is set as $x_2=\sqrt{x_1}$, and the
velocity of the leader is
$v_{L_{x_1}}(t)=0.02,~v_{L_{x_2}}(t)=\sqrt{0.02(t+1)}-\sqrt{0.02t}$.
In our simulations, the time label $t$ is a discrete number with
step length being equal to $1$. 

\begin{figure}[htp]
\centering \leavevmode
\begin{tabular}{cc}
\resizebox{4.2cm}{!}{\includegraphics[width=11.2cm]{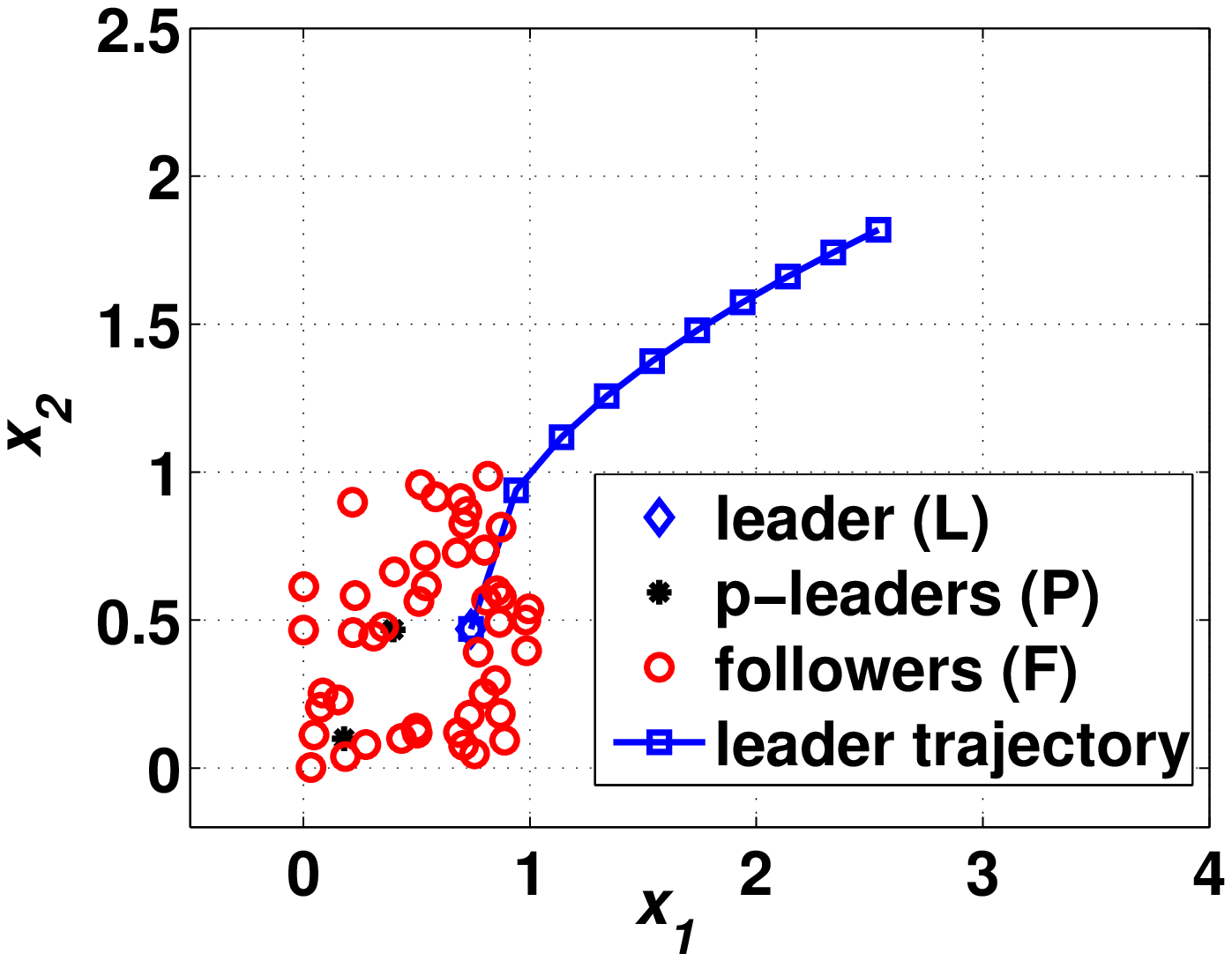}}
&
\resizebox{4.2cm}{!}{\includegraphics[width=11.2cm]{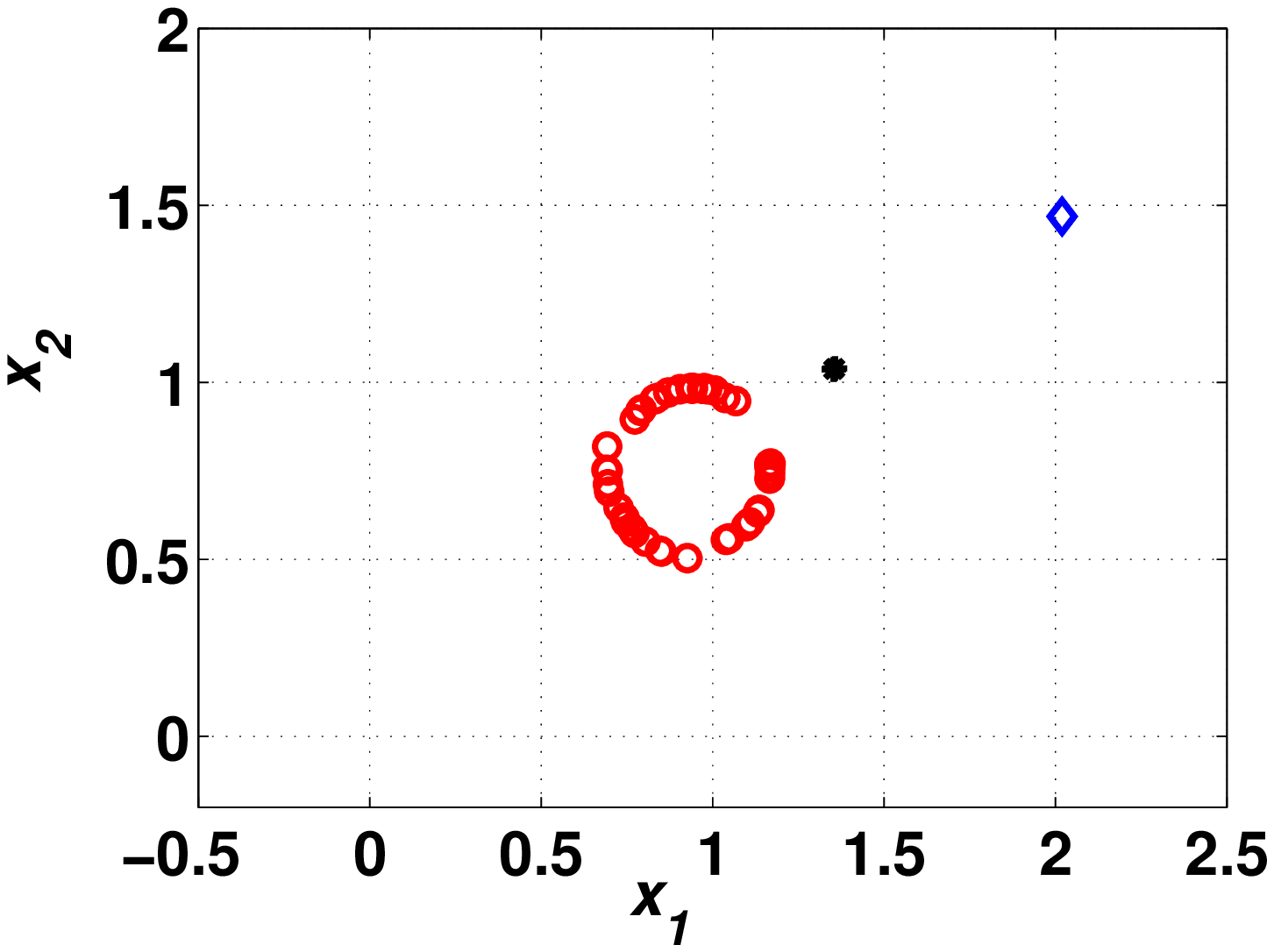}}
 \\
{\scriptsize (a) } &  {\scriptsize (b) } \\
\resizebox{4.3cm}{!}{\includegraphics[width=11.2cm]{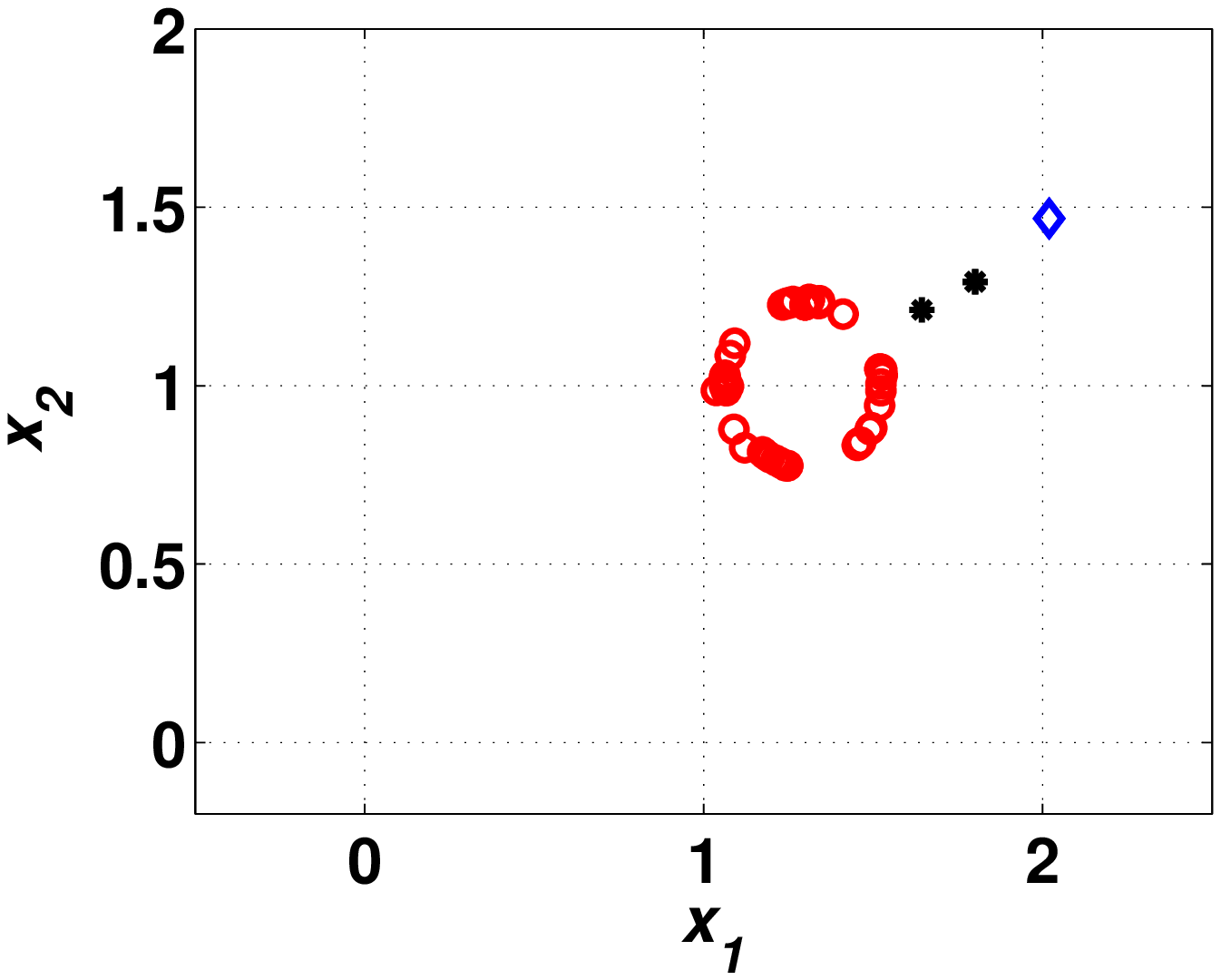}}
&
\resizebox{4.2cm}{!}{\includegraphics[width=11.2cm]{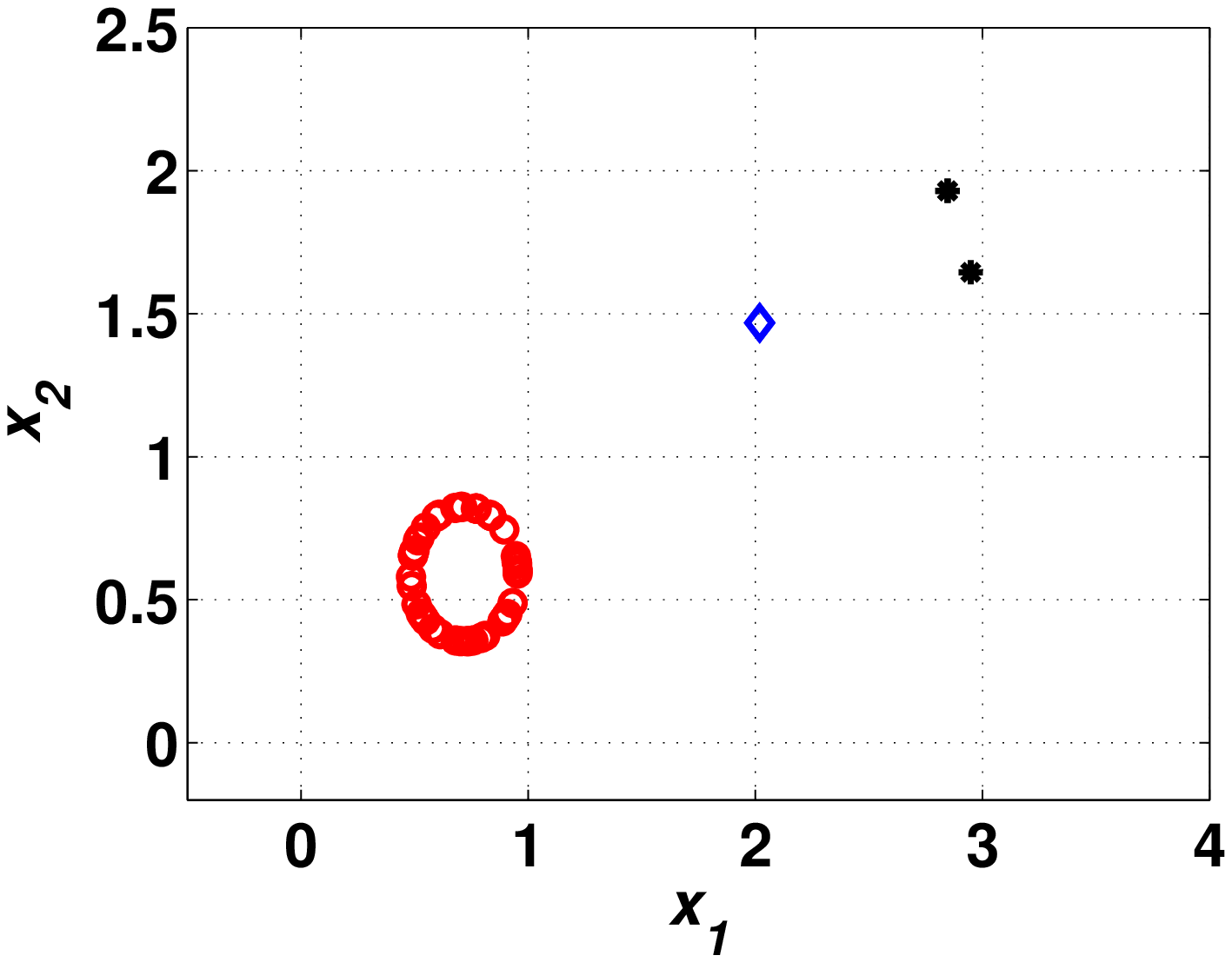}}\\
{\scriptsize (c) } &  {\scriptsize (d) }
\end{tabular}
\caption{(Color online) (a) Starting position of the 50-agent flock
consisting of 1 leader, 2 pseudo-leaders and 47 followers. The blue line
marked by square points denotes the trajectory of the leader. (b)
Flock position after 65 steps without predictive mechanism
($H_p=0$), (c) with proper predictive mechanism ($H_p=20$), and (d)
with over prediction ($H_p=70$). Here, $x_1$ and $x_2$ denote the two-dimensional position coordinates.} \label{fig:final position with
predictive mechanism}
\end{figure}

\begin{figure}[htp]
\centering \leavevmode
\begin{tabular}{cc}
\resizebox{4.3cm}{!}{\includegraphics[width=11.2cm]{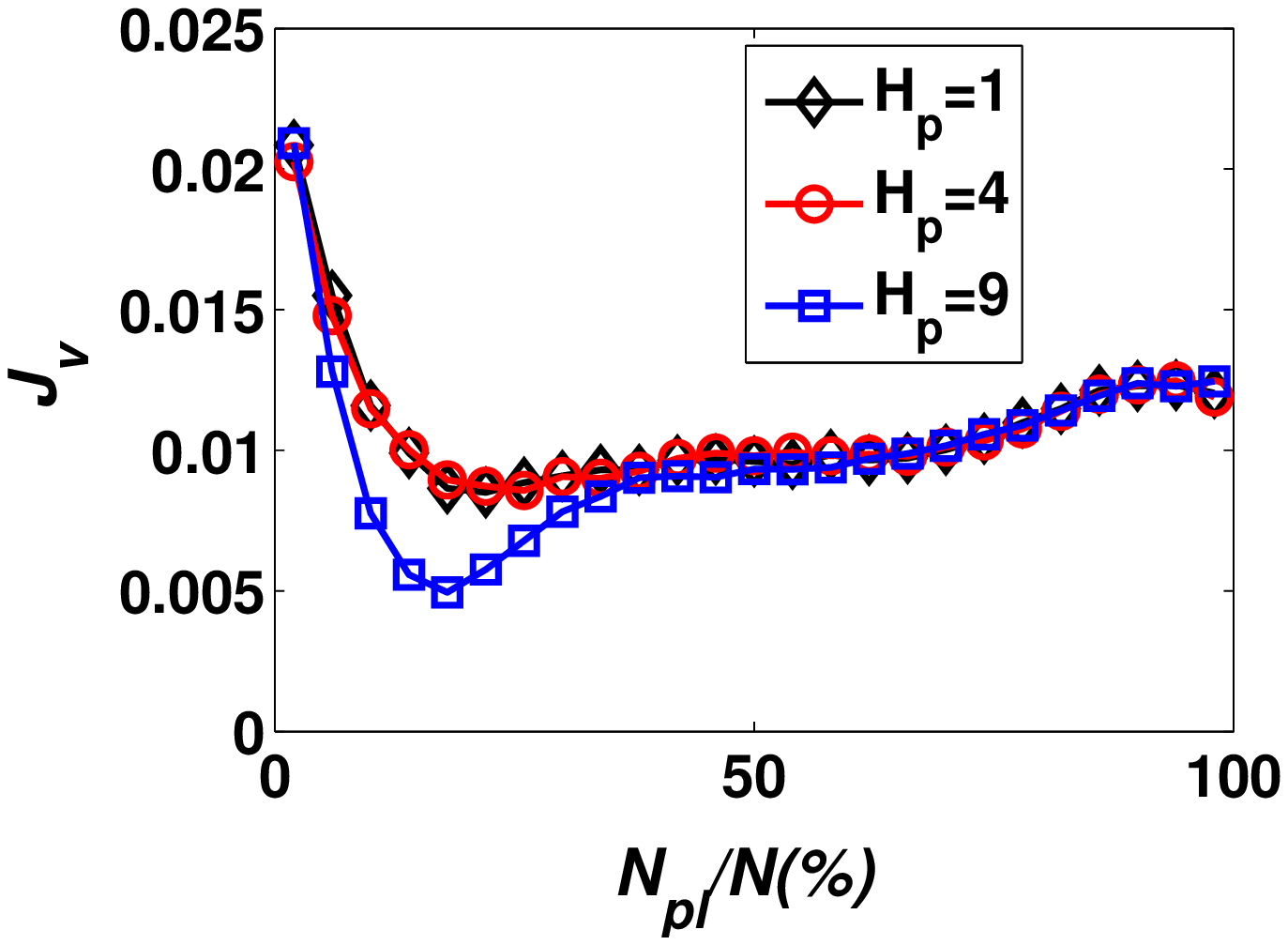}} &
\resizebox{4.3cm}{!}{\includegraphics[width=11.2cm]{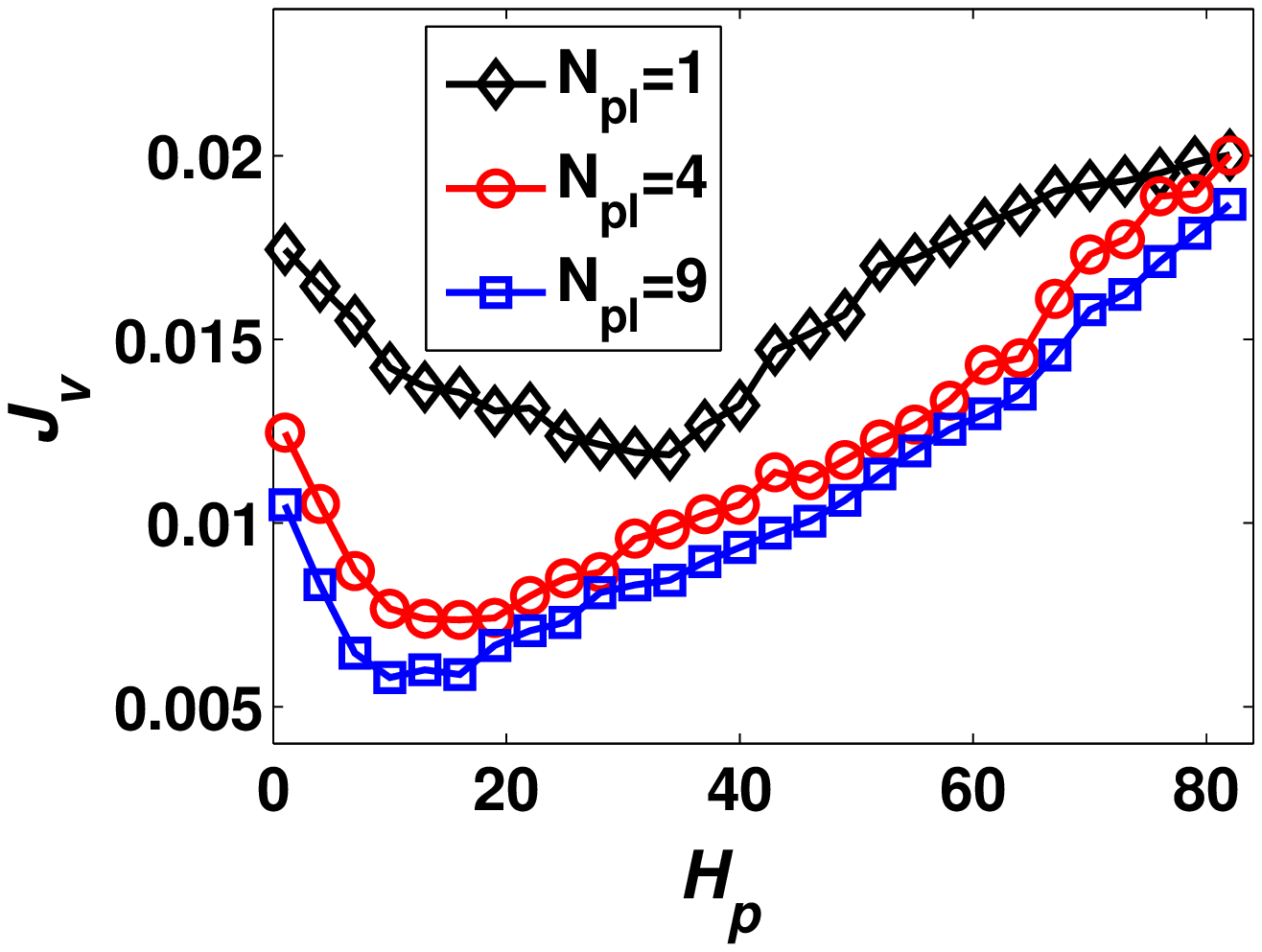}}
 \\
{\scriptsize (a) } &  {\scriptsize (b) } \\
\resizebox{4.3cm}{!}{\includegraphics[width=11.2cm]{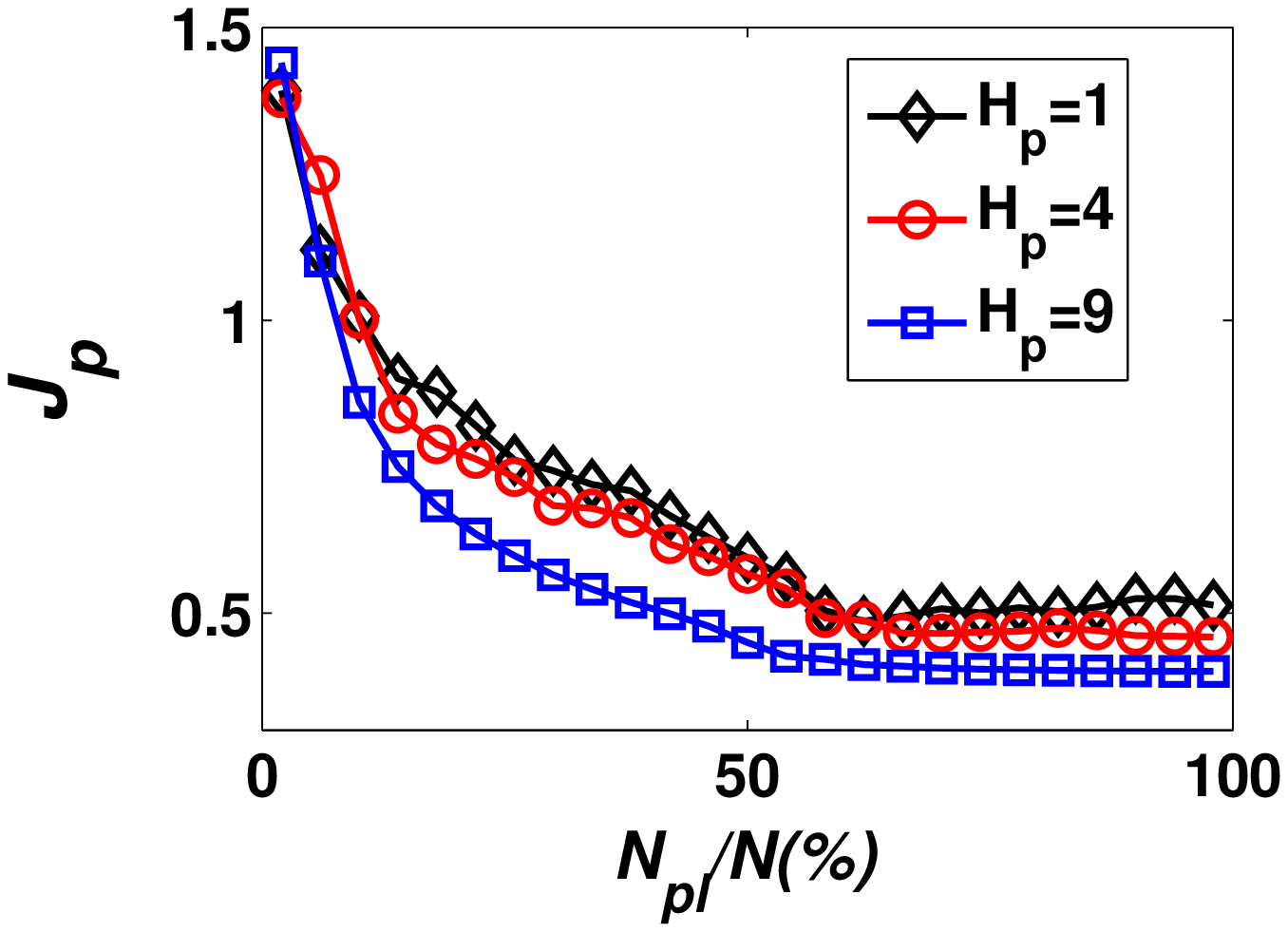}} &
\resizebox{4.3cm}{!}{\includegraphics[width=11.2cm]{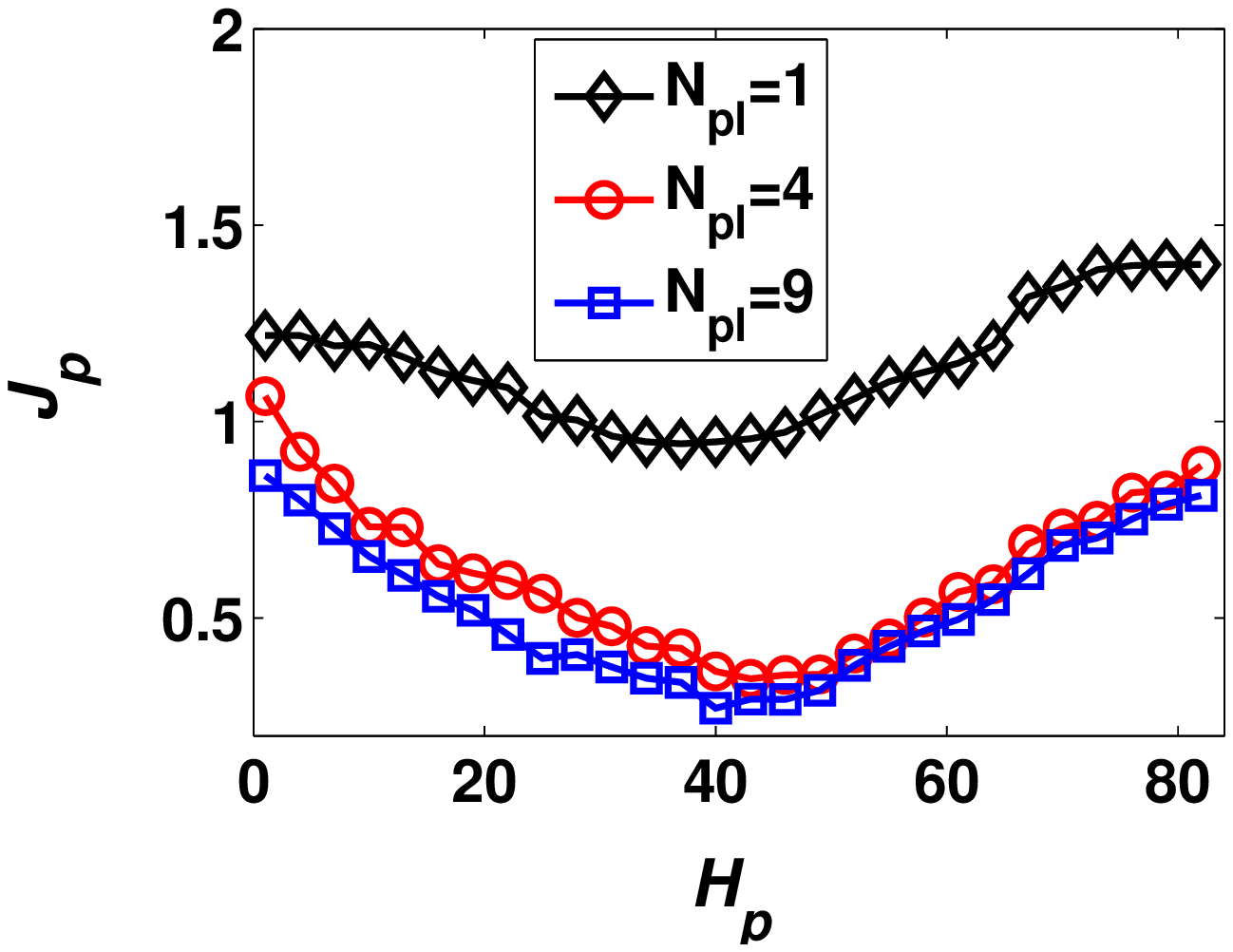}}\\
{\scriptsize (c) } &  {\scriptsize (d) }
\end{tabular}
\caption{(Color online) The roles of the pseudo-leaders' number $N_{pl}$ (figures (a) and (c))
and  prediction horizon $H_p$ (figures (b) and (d)) on a flock of $50$ agents.
The leader and the pseudo-leaders are selected randomly among these agents.
Each point is an average over $1000$ independent runs. The parameters
of the A/R functions~(\ref{eq: AR function}) and (\ref{eq:
AR function of followers}) are $a=8$, $b=17.6$, $c=0.4$, and $\tilde{a}=1$,
$\tilde{b}=2.2$, $\tilde{c}=0.2$, respectively. The radius of the influence
circle is $r=0.65$. Each agent starts from a
position randomly selected in the square $[0,1]\times [0,1]$.
Without loss of generality, the trajectory of the leader is
set along the curve defined by $x_2=\sqrt{x_1}$, and the velocity of the leader is
$v_{L_{x_1}}(t)=0.02,~v_{L_{x_2}}(t)=\sqrt{0.02(t+1)}-\sqrt{0.02t}$.} \label{fig:the
roles of Hp and pseudo-leaders}
\end{figure}

It can be seen from Figs.~\ref{fig:final position with predictive
mechanism}b--\ref{fig:final position with predictive mechanism}d
that, with a proper $H_p$, the coherence of the flock will be
improved remarkably. The followers will adhere to the leader much
more tightly (see Fig.~\ref{fig:final position with predictive
mechanism}c), and the flock formation will be more stable. More
precisely, for the flocks with the predictive mechanism, the
position error index
\begin{equation}
\label{eq: position error index} J_p=\frac{1}{N-1}\sum_{i=1,i\neq L
}^{N}\|d_{iL}\|_2
\end{equation}
converges to a constant after finite steps, indicating a stable
state of the flock dynamics. Here, $J_p$ measures the cohesion
performance of the flock, with $\|d_{iL}\|_2$ denoting the Euclidean
distance between agent $i$ (F or P) and the leader. Meanwhile, as to
the flock without this mechanism, as shown in Fig.~\ref{fig:final
position with predictive mechanism}b, $J_p$ will keep increasing
along with the elapse of time, making the flocking unstable.
However, abusing the foresight, namely over-prediction (see
Fig.~\ref{fig:final position with predictive mechanism}d), is also
undesirable. That is because the pseudo-leaders are
attracted/repelled by the leader position too many steps ahead and
will probably escape the flock with a fairly high speed, and then
lose influences on the followers. In this way,
the flocking will be damaged after finite steps.

The circular formation of the followers in
Figs.~\ref{fig:final position with predictive
mechanism}b--\ref{fig:final position with predictive mechanism}d is
because of the particular form of the A/R function. To be clear, we give a more detailed explanation as
follows: Actually, the biological flock dynamics is fairly complex.
For example, fish uses sidetrack to sense the current variances, in
this way, the fish schools are formed. On the other hand, in order
to save flying energy for the rear individuals, wild geese flocks
always form a `$\Lambda$'-like formation \cite{kr02}. In this case, the
forming process of such flock is determined by aerodynamics. Our
proposed model is based on the idea that an individual inside a
bio-group can predict the trajectory of its leader(s), and this kind of intelligence can help the individual
make more efficient decision to improve the flocking performance.  This paper, however,  does not aim at reproducing the detailed movement formations of any particular bio-groups. 

In order to extract the role of $H_p$ and the number of
pseudo-leaders denoted by $N_{pl}$, we display their influences on
the position error index $J_p$ and velocity error index $J_v$ in
Fig. 5, where
\begin{equation}
\label{eq: velocity error index} J_v=\frac{1}{N-1}\sum_{i=1, i\neq L
}^{N}\|\overrightarrow{v_i}-\overrightarrow{v_L}\|_2.
\end{equation}
Here, $J_v$ measures the formation performance of the flock, where
$\overrightarrow{v_L}$ and $\overrightarrow{v_i}$ denote the
velocity vectors of the leader and the $i$th agent (F or P). If
$J_v\rightarrow 0$, the relative velocity of each pair of agents approaches zero, thus the flock formation is fixed. In
Fig.~\ref{fig:the roles of Hp and pseudo-leaders}a, we fix $H_p$ and
display the curves of $J_v$ with increasing $N_{pl}$, while
Fig.~\ref{fig:the roles of Hp and pseudo-leaders}b, on the contrary,
reports the curves of $J_v$ with increasing $H_p$ and fixed
$N_{pl}$. It can be seen from Fig.~\ref{fig:the roles of Hp and
pseudo-leaders}a that the curves fall sharply at the beginning and
then more slowly until reaching a minimum, afterwards rising slowly
as the increasing of $N_{pl}$. It implies that adding just very few
pseudo-leaders (e.g. long-range links) to the leader, which
transforms the flock topology from a strongly localized network into
a small-world one, will improve the flocking performance greatly.
However, when the number of pseudo-leaders reaches an optimum
$N_{pl}^*$ corresponding to the minimal $J_v^*$, the flock formation
performance will start to worsen and these extra pseudo-leaders
become redundant. On the other hand, increasing $H_p$ can help
reduce $J_v$ in two ways: (i) it depresses $J_v$ with the same
$N_{pl}$; (ii) it reduces the optimal value of $N_{pl}^*$
corresponding to the minimal $J_v^*$. Compared with
Fig.~\ref{fig:the roles of Hp and pseudo-leaders}a, the $J_v$ curves
in Fig.~\ref{fig:the roles of Hp and pseudo-leaders}b fall more
slowly at the beginning until reaching the minimum, afterwards $J_v$
increases all along and never reaching a stable platform. It implies
that the flock formation performance can be remarkably improved with
proper predictive capability, however, too much vision into the
future, namely over-prediction, will even worsen the formation of
flocking. On the other hand, more pseudo-leaders, i.e., larger
$N_{pl}$
 (as long as $N_{pl}\leq N_{pl}^*$), can also help yield more
cohesive flocking with better formation.

\begin{figure}[htb]
\centering \leavevmode {\includegraphics[width=6.0cm]{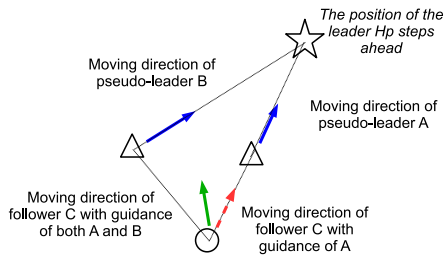}}
\caption{(Color online) Illustration of Fig.~\ref{fig:the roles of
Hp and pseudo-leaders}a. Follower C heads in the red dashed arrow
direction with the guide of pseudo-leader A who moves nearly in the
same direction as the leader, making the movement direction nearly
synchronized. However, the addition of pseudo-leader B will make the
direction of C deviate from the red dashed arrow to the green arrow, and thus
worsen the velocity synchronization performance $J_v$. Therefore,
more pseudo-leaders are not necessarily beneficial to synchronizing
the velocities of the flock.} \label{fig:pseudo-leader}
\end{figure}

An interesting phenomenon can be observed from Fig.~\ref{fig:the
roles of Hp and pseudo-leaders}a that increasing the number of
pseudo-leaders beyond some characteristic number makes the flocking
performance worse, which seems counter-intuitive. This can be
explained as follows. First,  as
shown in Eqs.~(\ref{eq: pseudo-leader dynamics}) and (\ref{eq:
follower dynamics}), the velocities of both the pseudo-leaders and
the followers are solely determined by the relative positions.
Indeed, as shown in Fig.~\ref{fig:pseudo-leader}, if there is only
one pseudo-leader A, then the follower C will head in the red dashed arrow
towards A, and all the three individuals 
are nearly moving in the same direction, which is desirable.
However, when another pseudo-leader B is added into this flock, the
follower C will move towards a certain position between A and B (see
the green arrow). Apparently, the velocity synchronization
performance is worsened compared with the former case. Therefore, the
observation in Fig.~\ref{fig:the roles of Hp and pseudo-leaders}a is
reasonable. To give a more vivid and detailed explanation, we give further discussions 
in \textbf{Appendix A} for interested readers.

Next, we investigate the effects of $H_p$ and $N_{pl}$ for another
important index $J_p$. It can be seen from
Fig.~\ref{fig:the roles of Hp and pseudo-leaders}c that the curves
fall sharply at the beginning and then asymptotically approach a
stable value. The main difference between Fig.~\ref{fig:the roles of
Hp and pseudo-leaders}a and Fig.~\ref{fig:the roles of Hp and
pseudo-leaders}c is that the latter is monotonous and has no
minimum, in other words, the increase of $N_{pl}$ always improves
$J_p$. A special case explaining this phenomenon is that if all the
followers serve as the pseudo-leaders, then they will be very
cohesive to the leader. However, when $N_{pl}$ exceeds a certain
value $\bar{N}_{pl}$, $J_p$ will increase so slowly that almost no
substantial improvement can be achieved. Moreover, increasing $H_p$
can help reduce $J_p$, and the improvement shrinks along with
increasing $H_p$, indicating a saturation effect that no remarkable
improvement can be achieved by pushing $H_p$ to a very high value
(which may cost too much). Compared with Fig.~\ref{fig:the roles of
Hp and pseudo-leaders}c, the curves of Fig.~\ref{fig:the roles of Hp
and pseudo-leaders}d decrease more slowly first and then reach the
lowest values at a fairly large $H_p$, afterwards rise quickly.
Thus, over-prediction is not preferred. Analogous to
Fig.~\ref{fig:the roles of Hp and pseudo-leaders}b, more
pseudo-leaders will help improve the cohesive flocking performance,
however, this improvement also displays a saturation effect with
increasing $N_{pl}$. Actually, when $N_{pl}$ exceeds a certain
value, $J_p$ increases very slowly that almost no benefit can be
gained by further increasing $N_{pl}$. In brief, suitable insight
into the future and moderate number of pseudo-leaders are preferred.

A plausible physical rule behind the observed phenomena shown in
Fig.~\ref{fig:the roles of Hp and pseudo-leaders} is that, in order
to achieve a fixed flocking performance, greater predictive
capability and more pseudo-leaders can compensate the insufficiency
of each other. Given a fixed-size flock, separately for formation
performance $J_v$ and cohesive performance $J_p$, there exists an
optimized combination of $H_p$ and $N_{pl}$. The conclusions
achieved in this paper are not sensitive to the trajectory of the
leader. In order to validate the generality of the conclusions, we have used another two
different trajectories, i.e., parabolic and sinusoidal curves. The
simulation results are shown in \textbf{Appendix B}, which suggest
that our main conclusion, i.e., predictive capability and long-range
links can compensate for the insufficiency of each other, also
holds in those two cases.

\begin{figure}[htp]
\centering
\begin{tabular}{cc}

{\includegraphics[width=4.5cm]{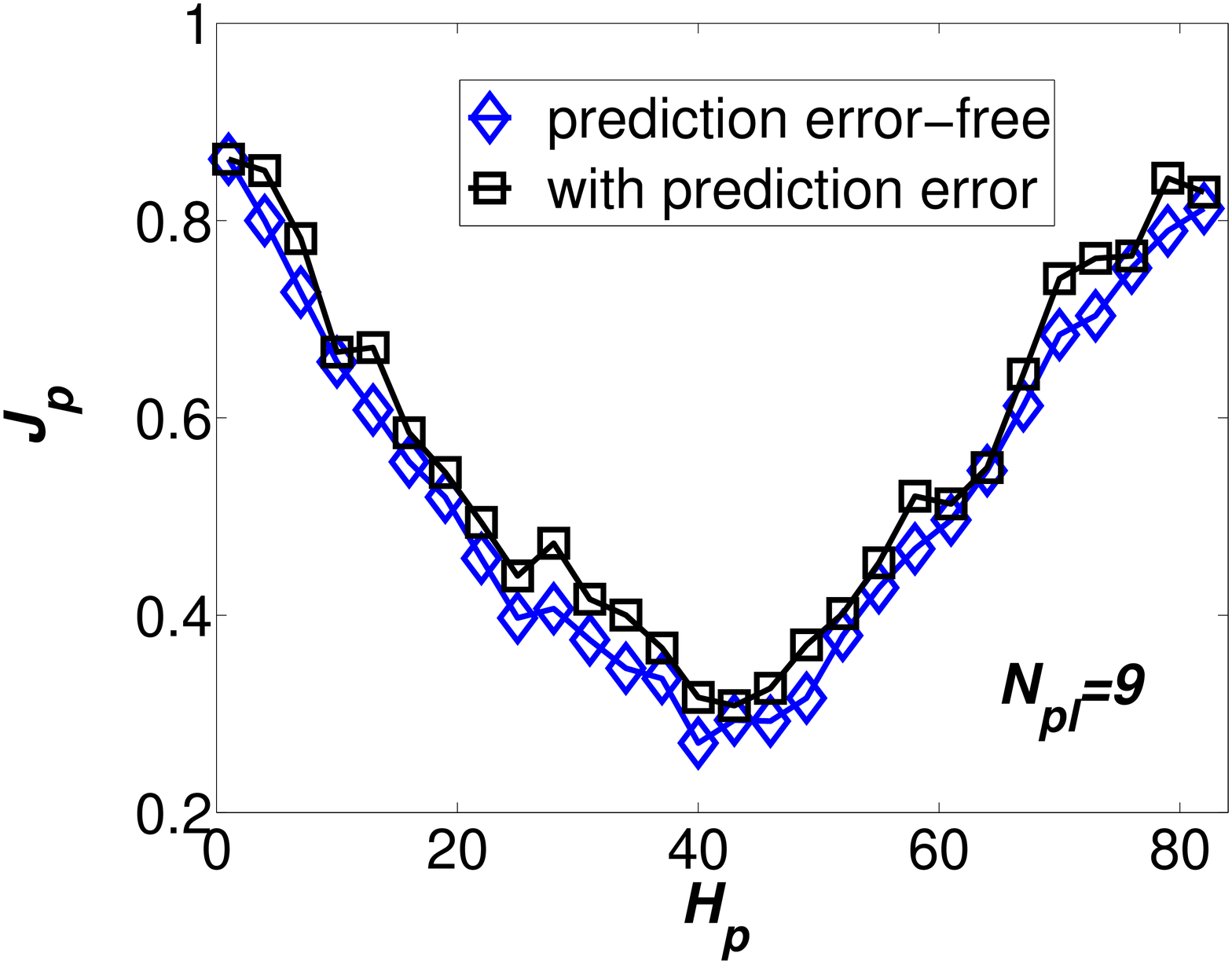}}&
{\includegraphics[width=4.5cm]{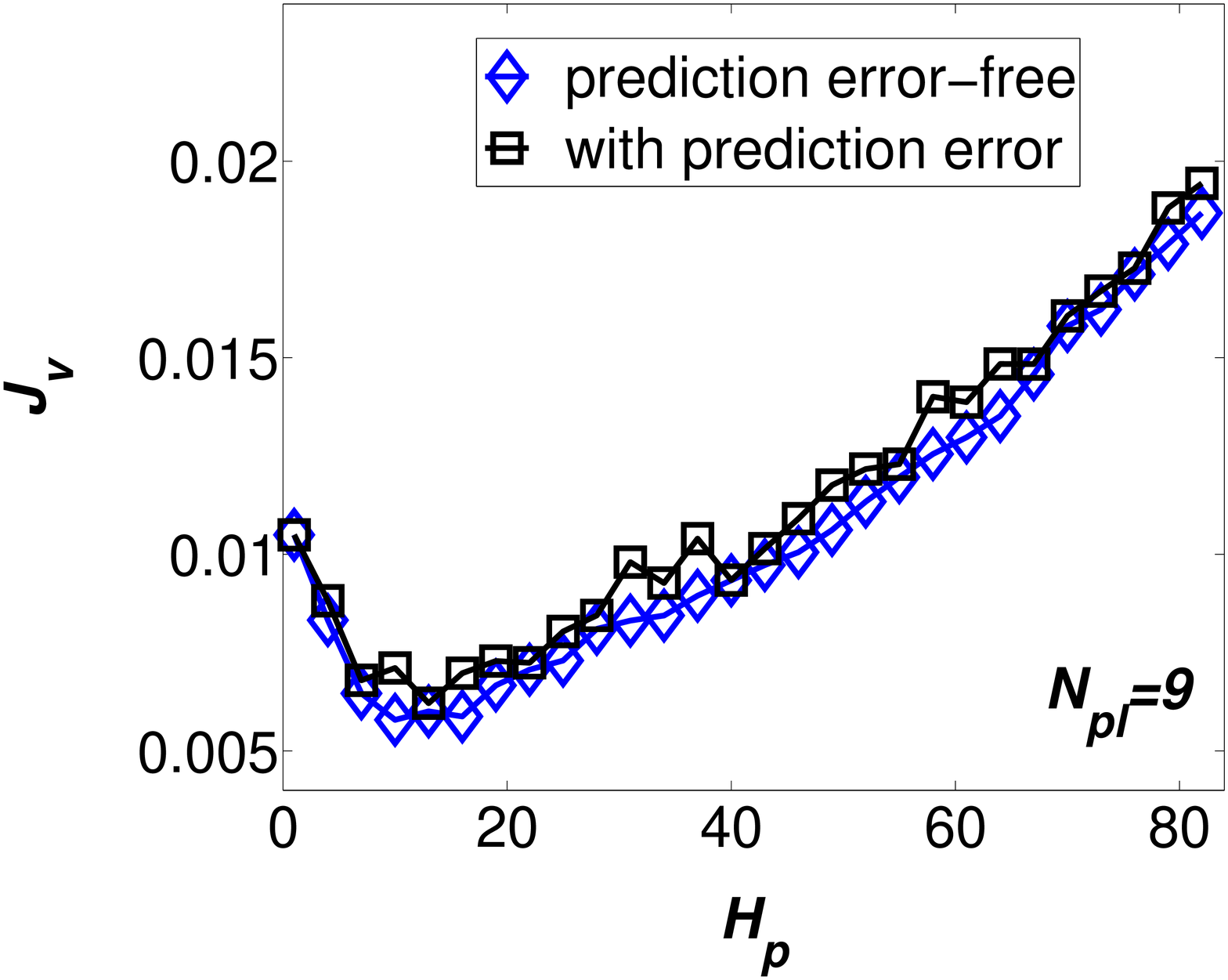}}
%
 \\
{\scriptsize (a)} &  {\scriptsize (b)}
\end{tabular}
\caption{(Color online) With-noise case vs. noise-free case for a
flock with $N=50$, $N_{pl}=9$, and $\eta=0.02$. The other parameters
and the initial conditions are the same as those in
Fig.~\ref{fig:the roles of Hp and pseudo-leaders}. Each point is an
average over 1000 independent runs. The moderate external noise does
not change the global behavior of the flock.}
 \label{fig: A/R with noise}
\end{figure}

From the practical point of view, external perturbations (noise) and
internal modeling mismatch are always present in any realistic
systems, which inevitably induces some prediction error for the
pseudo-leaders. To examine the influence of such prediction errors,
we now introduce the perturbation into Eqs. \eqref{eq: AR function}
and \eqref{eq: AR function of followers} by adding an external
two-dimensional white noise term $\xi \in
[-\frac{1}{2}\eta,\frac{1}{2}\eta]\times
[-\frac{1}{2}\eta,\frac{1}{2}\eta]$ to the vector $d_{pL}$, i.e.,
$\hat{d}_{pL}={d}_{pL}+\xi$ and
$G(\hat{d}_{pL})=-\hat{d}_{pL}\left(a-b\cdot
\mbox{exp}(-\|\hat{d}_{pL}\|_2^2/c)\right)$. The leader's position
is no longer perfectly known to the pseudo-leaders. From
Fig.~\ref{fig: A/R with noise}, one can observe that moderate
external noises do not change the global behavior of the flock. The
tendency of the curves $J_p$ and $J_v$ are almost the same as the noise-free case. Furthermore, to understand the capacity and
robustness of our purposed predictive mechanism more deeply, we have
also investigated the influences of stronger prediction errors $\xi$
with other kinds of leader trajectories including parabolic and
sinusoidal curves in \textbf{Appendix C}. We found that the
tolerance range of prediction error is large enough. In this way,
the generality of the conclusions on the role of the predictive
mechanism is thus further verified.

\section{Predictive mechanisms in the Vicsek model}\label{Sec: prediction Vicsek}

The role of predictive mechanisms highlighted in
Section~\ref{Sec: AR results} is not merely confined to A/R flocks but
quite general. To verify this, we now incorporate this
predictive mechanism into another flocking model,
i.e., the Vicsek model \cite{vi95}, and compare the synchronization
performance of the predictive small-world Vicsek model with the one of the standard Vicsek model.

\begin{figure}[htp]
\centering
\begin{tabular}{c}
\hspace*{-0.8cm}
\resizebox{7cm}{!}{\includegraphics[width=11.2cm]{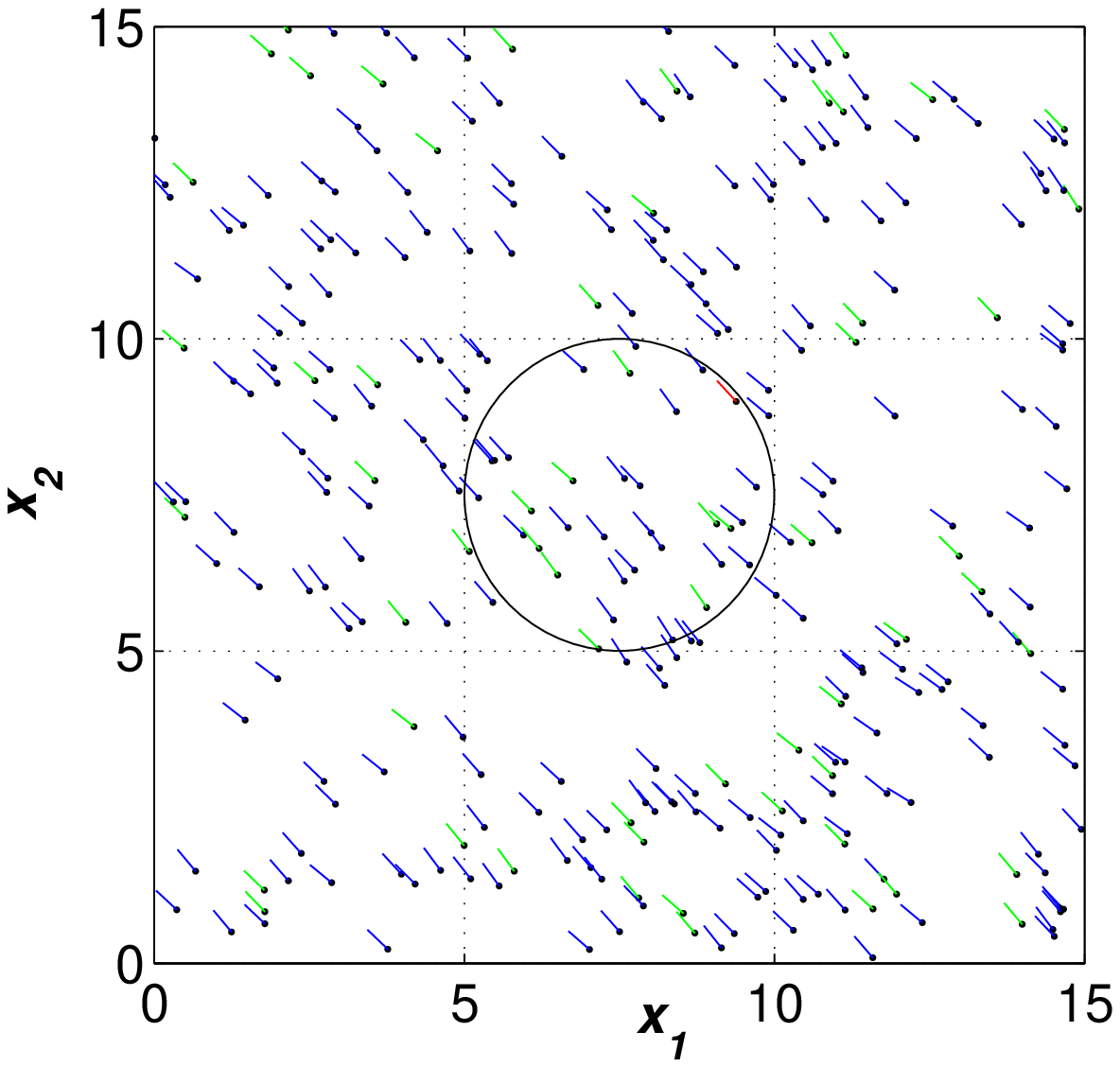}}
\\ {\scriptsize (a)}\\\hspace*{-0.6cm}
\resizebox{7cm}{!}{\includegraphics[width=11.2cm]{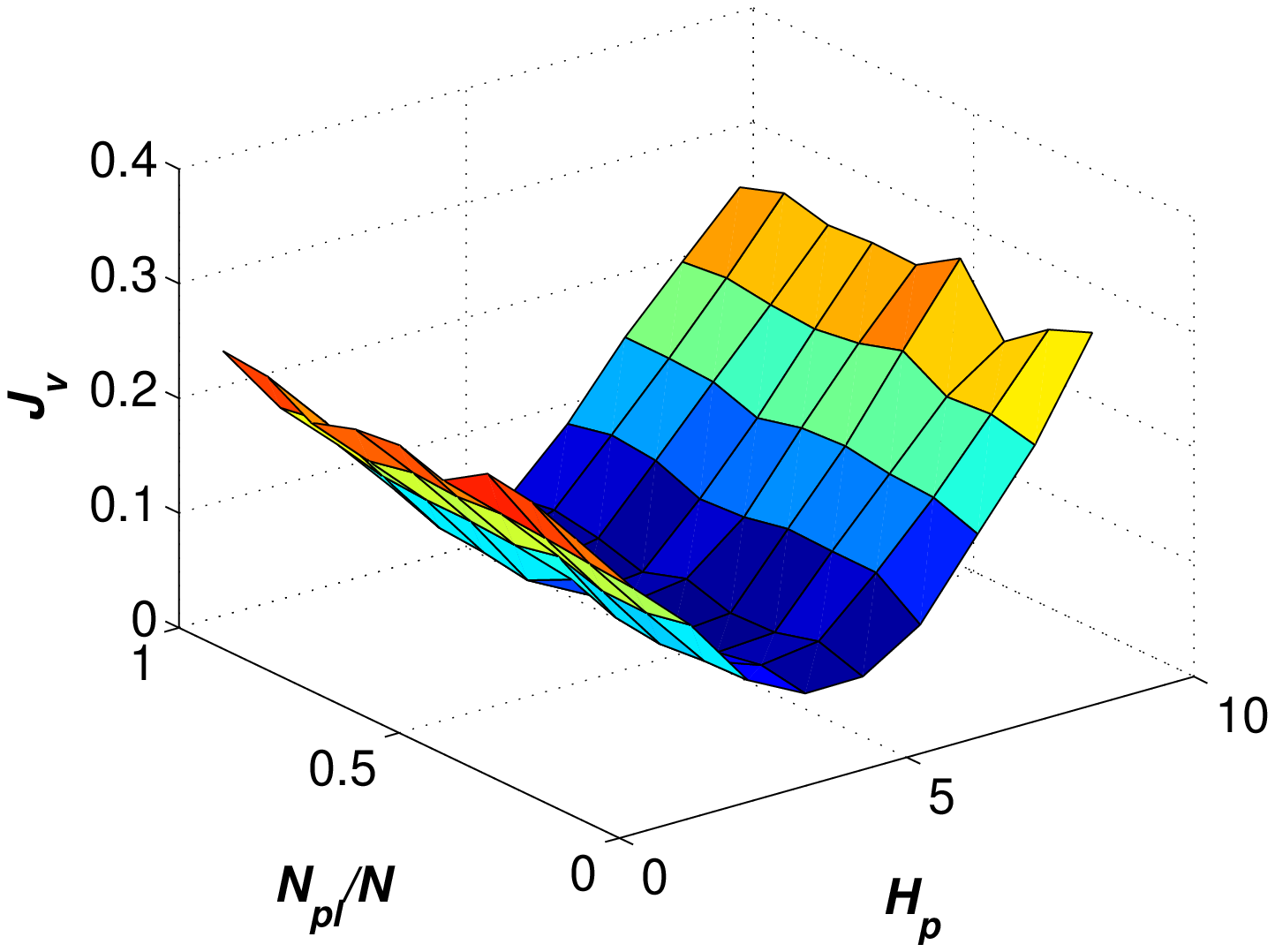}}
 \\
{\scriptsize (b) }
\end{tabular}
\caption{(Color online) (a) Snapshot of the predictive Vicsek flock
at the $12^{th}$ running step. The red particle denotes the leader;
the green particles represent the pseudo-leaders and the blue
particles denote followers. The centered black circle outlines the
trajectory of the leader. In this case, the prediction
horizon is $H_p=4$. (b) Velocity synchronization index $J_v$ as a
function of the parameters $H_p$ and $N_{pl}$. The parameters of
this simulation are $L=15$, $\eta=0.1$, $v=0.15$,
$N=300$, $r=1$, and $R=L/6$. The simulation result is averaged over $1000$
independent runs.}
 \label{fig: the role of predictive mechanism}
\end{figure}

In this model, the velocities $v_i$ of the $N$ agents composing the group
are determined simultaneously at each discrete-time instant, and the
position of the $i$th agent is updated according to
\[ x_i (k+1)=x_i(k)+v_i(k),\]
where $v_i(k)$ denotes the velocity vector of agent $i$ at time $k$.
For each agent the velocity vector, $v_i(k)$, is characterized by a constant magnitude $v$ and by a
direction $\theta_i(k)$ whose dynamics is given by
\[\theta_i (k+1)=\left<\theta_i\left(k\right)\right>_r+\Delta\theta_i,\]
where $\left<\theta_i\left(k\right)\right>_r$ denotes the average
direction of all the agents' velocity vectors within a circle of
radius $r$ centered on agent $i$, i.e.,
\[\left<\theta_i\left(k\right)\right>_r=\left\{\begin{array}{l}\mbox{arctan}\left[\left<\sin\left(\theta_i\left(k\right)\right)\right>_r/
\left<\mbox{cos}\left(\theta_i\left(k\right)\right)\right>_r\right]\\
\quad \quad \mbox{if}~
\left<\mbox{cos}\left(\theta_i\left(k\right)\right)\right>_r\geq
0;\\\mbox{arctan}\left[\left<\sin\left(\theta_i\left(k\right)\right)\right>_r/
\left<\mbox{cos}\left(\theta_i\left(k\right)\right)\right>_r\right]+\pi\\
\quad \quad \mbox{otherwise},
\end{array}\right.\]
where $\left<\sin\left(\theta_i\left(k\right)\right)\right>_r$ and
$\left<\mbox{cos}\left(\theta_i\left(k\right)\right)\right>_r$
denote the average sine and cosine values, and $\Delta\theta_i$
represents a random noise obeying a uniform distribution in the
interval $[-\eta/2, \eta/2]$.

As shown in Fig.~\ref{fig: the role of predictive mechanism}a the
particles are distributed in a square of dimension $[0,L]\times
[0,L]$. The trajectory of the leader, which is not affected by others, is a circle centered at
$(L/2,L/2)$ with radius $R=L/6$ so that the direction of the leader
changes constantly. The small-world predictive connection framework
shown in Fig.~\ref{fig:structure} is used together with the Vicsek
model. Hence, the $N_{pl}$ pseudo-leaders  are always influenced by the leader's velocity $H_p$ steps ahead together with its neighbors' current velocities. It is shown in Fig.~\ref{fig: the role of
predictive mechanism}b that drastic improvement of the velocity
synchronization performance can be achieved with moderate prediction
horizons. Similar to the results of the A/R model shown in
Section~\ref{Sec: AR results}, one can also conclude that suitable
insight into the future and moderate number of pseudo-leaders is
preferable.

\section{Conclusion and discussion}\label{sec: conclusion}

Inspired by the predictive mechanisms that universally exist in
abundant natural bio-groups, we incorporate a certain predictive protocol
into flocks with small-world structure. The predictive mechanism
embedded in the pseudo-leaders lets many 
neighboring-connected followers know the current or even future
dynamics of the leader in time, thus each individual can make a
decision based on  timely information of the leader instead of the
delayed information as in some traditional models. In this way, the
followers become more cohesive to the leader and the flock formation
becomes more stable. Note that, in our model, the leader's motion
governs the trajectory of the whole swarm. However, it is a general
feature of real migration flocks. Actually, in \cite{co05}, a
changeable target known by a few leaders is used to guide the whole
flock, in which only the target's motion drives swarming behavior.
Analogously, in this paper, a single leader is used to determine the
general trajectory of the whole swarm. Therefore, the current
leader-driven swarm is not unrealistic.

Simulation results led to the following conclusions: (i) Increasing
the number of pseudo-leaders can always improve the cohesive
flocking performance. Furthermore, it can improve the formation
flocking performance when the pseudo-leader number has not exceeded
a threshold, otherwise, the performance will be degraded. (ii) There
exists a certain value of $H_p$ that optimizes the cohesive and the
formation flocking performances, in other words, moderately
increasing $H_p$ will improve the flocking performance, whereas
predicting too many steps ahead will impair the flocking. (iii)
Predictive capability and long-range links can compensate for the
insufficiency of each other. It is worthwhile to emphasize the
observed over-prediction phenomenon, which is of significance in
practice.

Furthermore, to verify the generality of these conclusions, we have
also applied the predictive mechanism to another popular flock
model, the directed graph
model with linear dynamics \cite{zh07,zh08}. The corresponding results
also strongly suggest that predictive protocols are beneficial to
flocking dynamics when taking into consideration both the flocking
performance and the communication cost. More importantly, with this
mechanism, only a very small proportion of the followers are
required to act as the pseudo-leaders to achieve a better flocking
performance, as measured by $J_p$ and $J_v$. From the industrial
application point of view, the value of this work is two-fold: (i)
The flocking performance is significantly improved by injection of a
suitable predictive mechanism into the pseudo-leaders; (ii)
Moderately increasing the predictive capability can help remarkably
decrease the required number of pseudo-leaders. The latter feature
is fairly useful for networks with insufficient long-range
communication links, which are routinely costly.

This work provides a starting point aimed at achieving better
flocking performance by using a predictive mechanism, and we hope
that it will open new avenues in this fascinating direction.

\begin{acknowledgements}
The authors acknowledge Prof. Guanrong Chen and Prof. Jan M.
Maciejowski for their valuable and constructive
suggestions. H.T.Z. acknowledges the support of National Natural
Science Foundation of China (NNSFC) under Grant No. 60704041, and
the Research Fund for the Doctoral Program of Higher Education
(RFDP) under Grant No. 20070487090. T.Z. acknowledges the support of
NNSFC under Grant No. 10635040.
\end{acknowledgements}

\section*{Appendix A: Supplemental illustration of Fig.~\ref{fig:the
roles of Hp and pseudo-leaders}a}

\begin{figure}[htb]
\centering \leavevmode
\begin{tabular}{c}
\resizebox{6.5cm}{!}{\includegraphics[width=11.2cm]{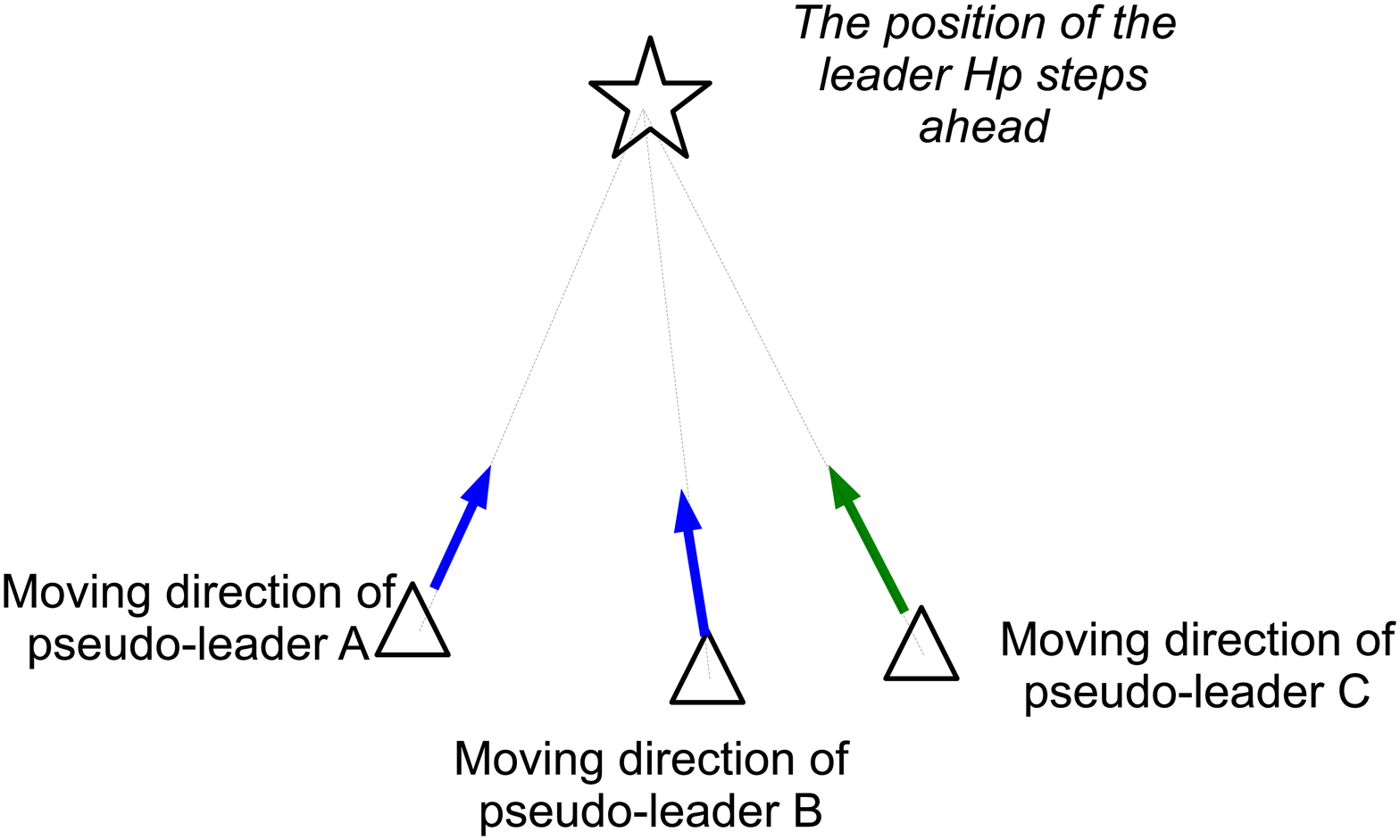}}\\
{\scriptsize (a) }\\
\resizebox{6.0cm}{!}{\includegraphics[width=11.2cm]{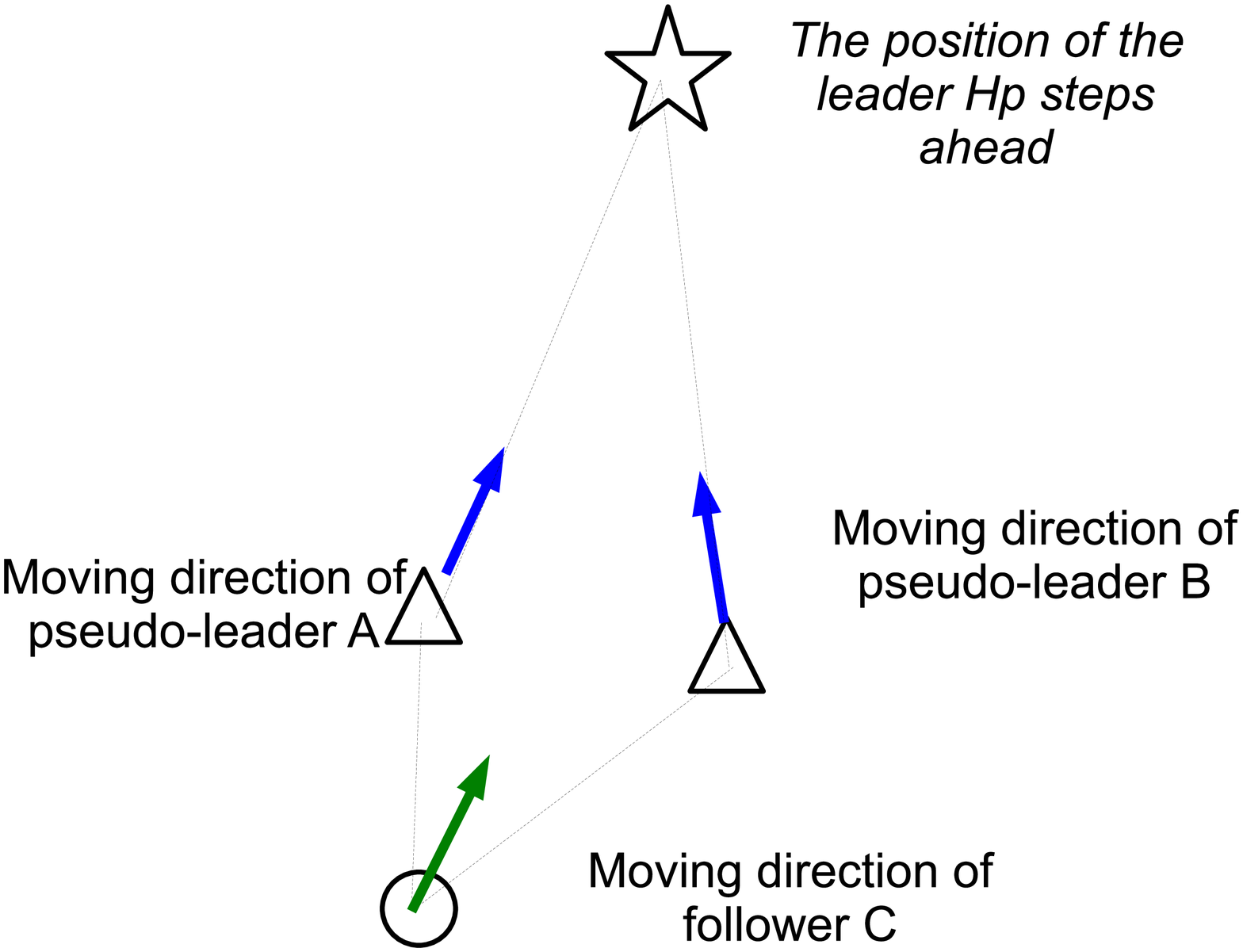}}
 \\
{\scriptsize (b) }
\end{tabular} \caption{Supplemental illustration of Fig.~\ref{fig:the
roles of Hp and pseudo-leaders}a. In this scenario, pseudo-leader A
moves nearly in the same direction of the leader, and degrading
pseudo-leader C (see sub-figure (a)) into a follower (see sub-figure
(b)) helps improve the velocity synchronization performance $J_v$.}
\label{fig:pseudo-leader-sub}
\end{figure}

We take a 4-agent flock for instance. As shown
in Fig.~\ref{fig:pseudo-leader-sub}a, all the individuals except the
leader are pseudo-leaders. However, due to the repulsion between
each couple of pseudo-leaders and the powerful attraction from the
leader, the pseudo-leaders will keep certain distances between each
other and form a section-like shape. Since all the pseudo-leaders
always point to the leader's predictive position $H_p$ steps ahead,
the moving directions of all the individuals are obviously far from
synchronized. By contrast, when one pseudo-leader C has been
degraded into a follower as shown in Fig.~\ref{fig:pseudo-leader-sub}b,
and take into consideration that the followers are always lagging
behind the pseudo-leaders, the direction of follower C is aligned to
approach the heading of the leader more or less. In this way, the
velocity synchronization performance $J_v$ is improved. Thus, too
many pseudo-leaders are not preferred.
For more general cases with prediction errors, more
long-range connections will introduce larger deviation, which
partially neutralize the positive effect of pseudo-leaders on $J_v$,
thus moderate pseudo-leaders are also desirable. 

\section*{Appendix B: Effects of the predictive mechanism with different leader's trajectories}
\begin{figure}[htp]
\centering \leavevmode
\begin{tabular}{cc}
\resizebox{4.3cm}{!}{\includegraphics[width=11.2cm]{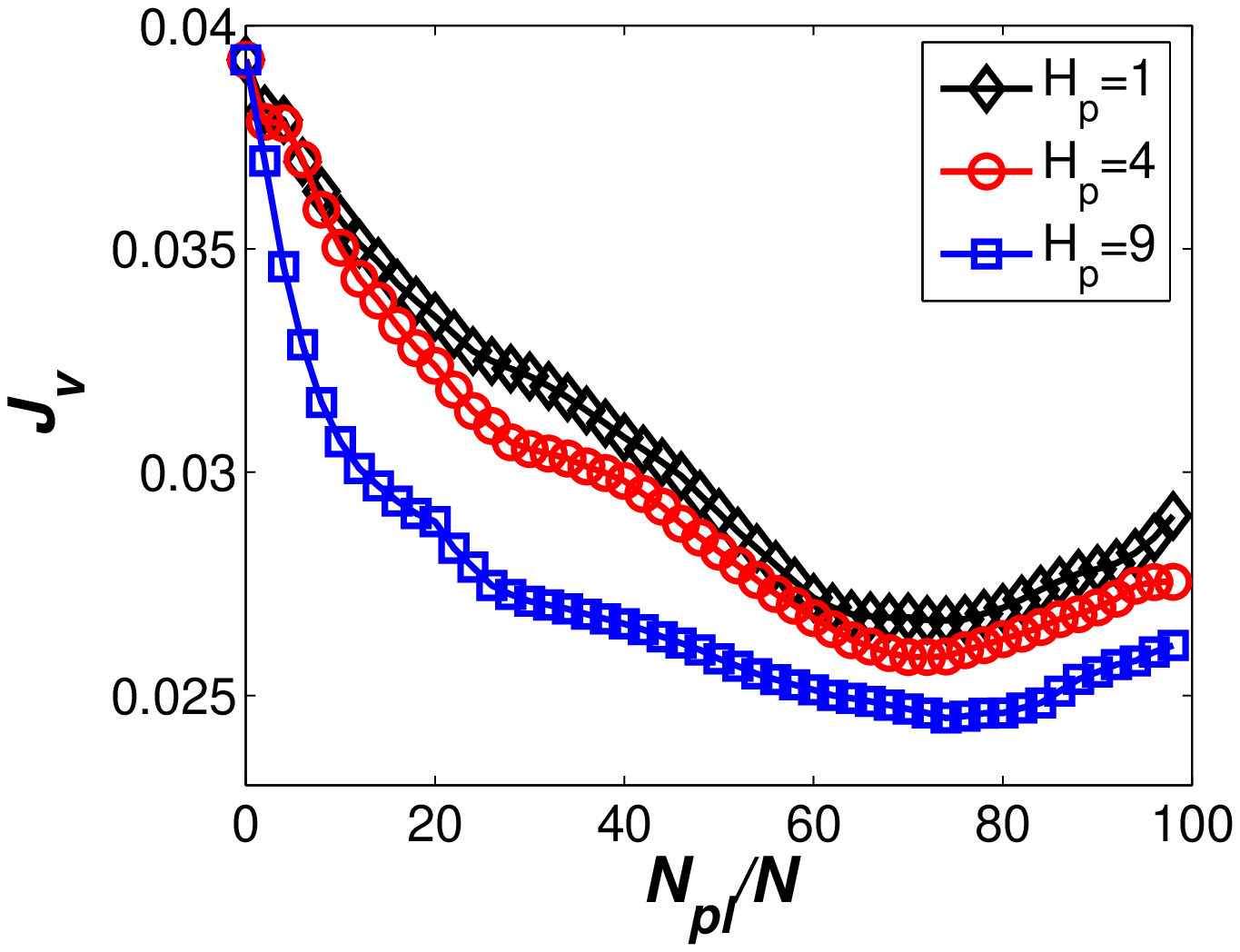}} &
\resizebox{4.3cm}{!}{\includegraphics[width=11.2cm]{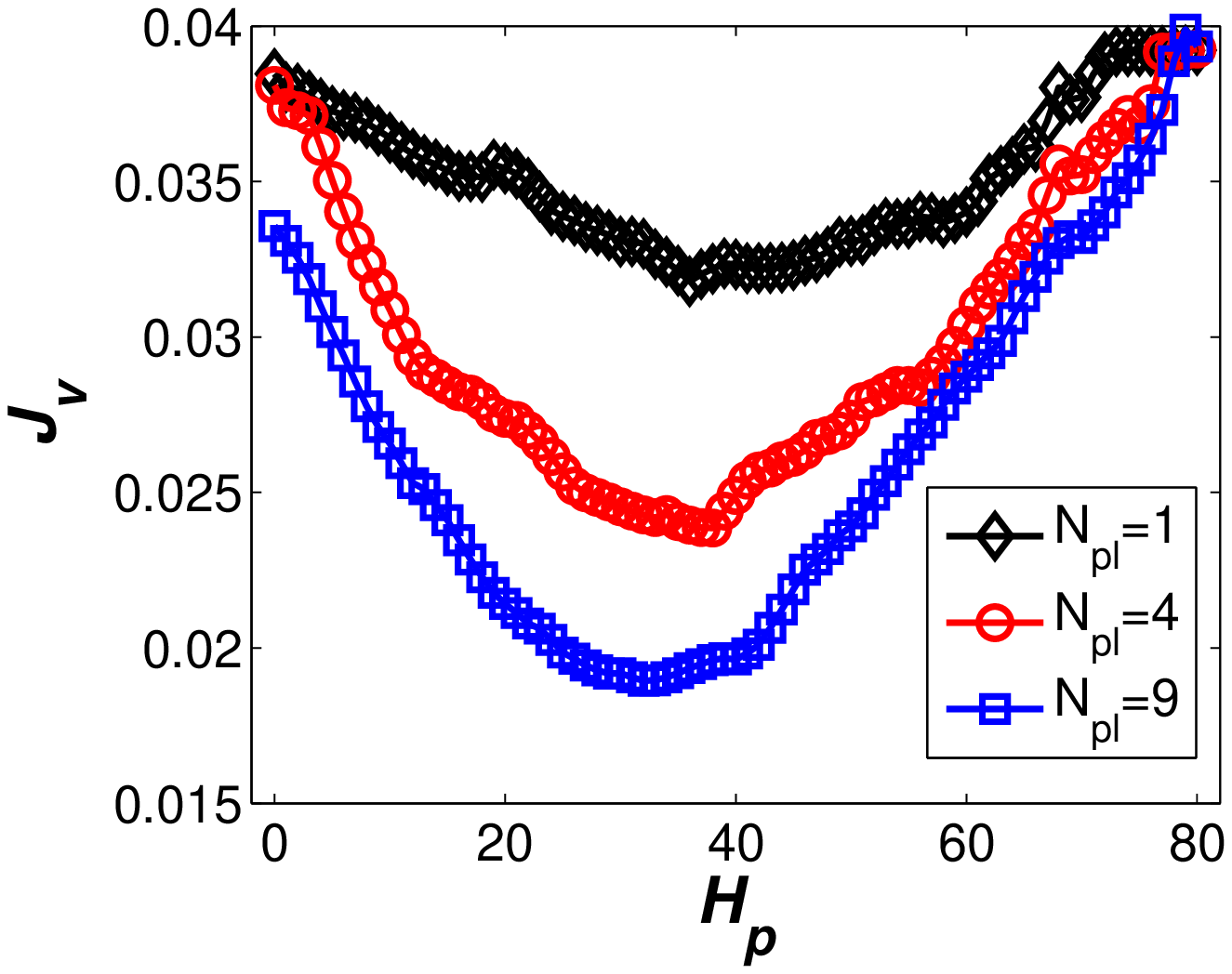}}
 \\
{\scriptsize (a) } &  {\scriptsize (b) } \\
\resizebox{4.3cm}{!}{\includegraphics[width=11.2cm]{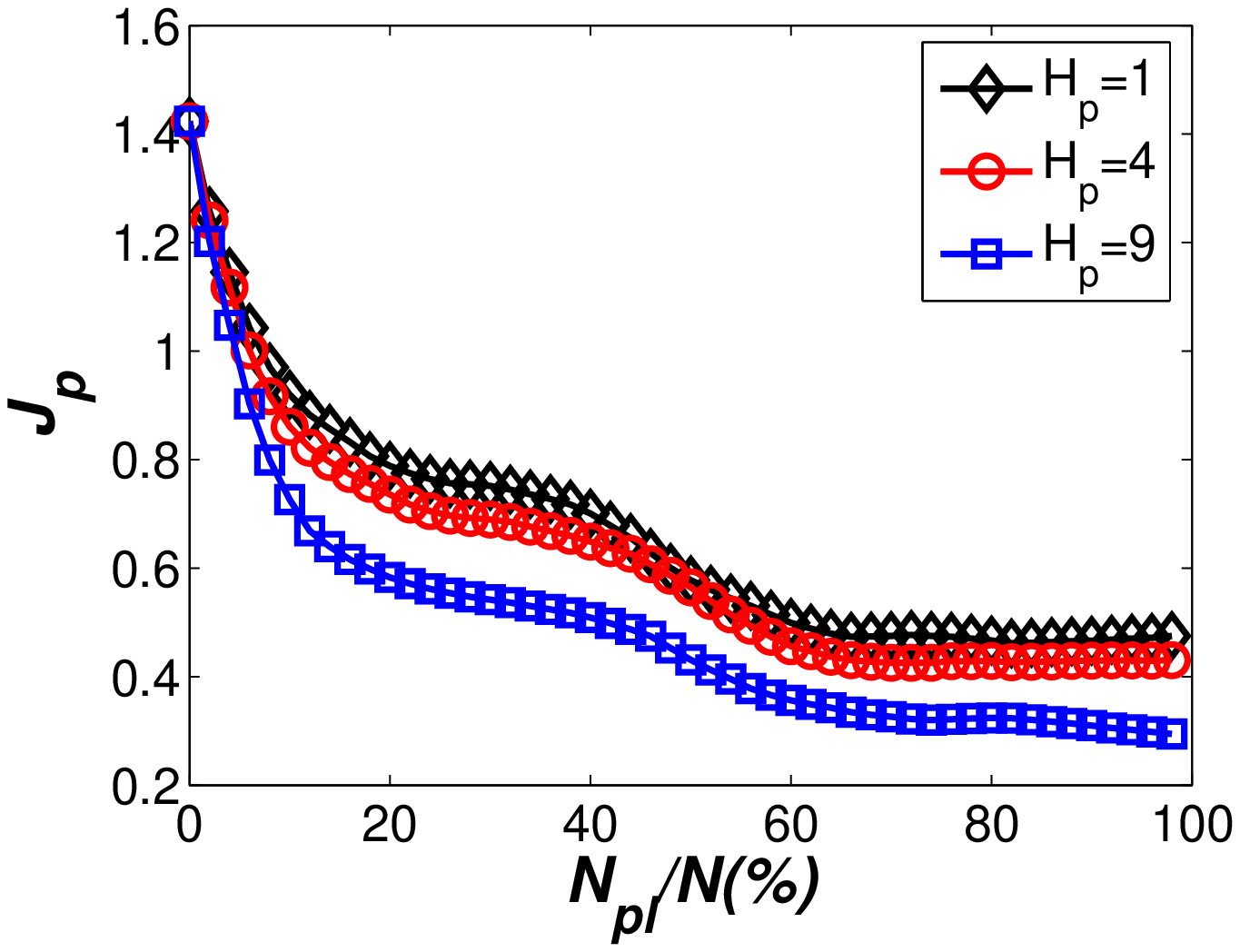}} &
\resizebox{4.3cm}{!}{\includegraphics[width=11.2cm]{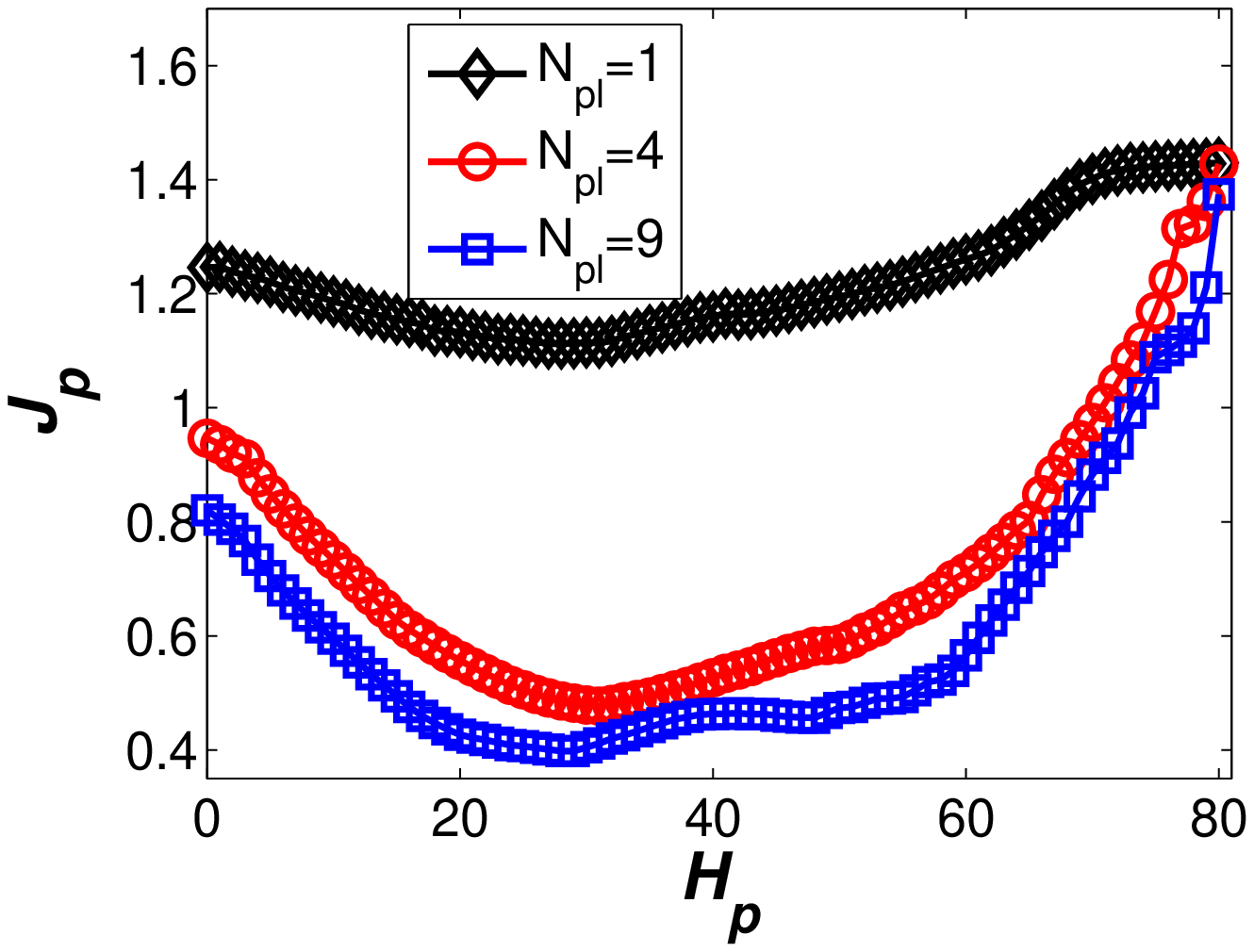}}\\
{\scriptsize (c) } &  {\scriptsize (d) }
\end{tabular}
\caption{(Color online) The roles of the pseudo-leaders' number
$N_{pl}$ (figures (a) and (c)) and prediction horizon $H_p$ (figures
(b) and (d)) on a flock with a total of $N=50$ agents. The
trajectory of the leader is set along the sinusoidal curve defined
by $x_2=\sin(2x_1)+1$, and the velocity of the leader is
$v_{L_{x_1}}(t)=0.02,~v_{L_{x_2}}(t)=\sin(0.04(t+1))-\sin(0.04t)$.
The other parameters and initial conditions are the same as those in
Fig.~\ref{fig:the roles of Hp and pseudo-leaders}.}
\end{figure}

\begin{figure}[htp]
\centering \leavevmode
\begin{tabular}{cc}
\resizebox{4.3cm}{!}{\includegraphics[width=11.2cm]{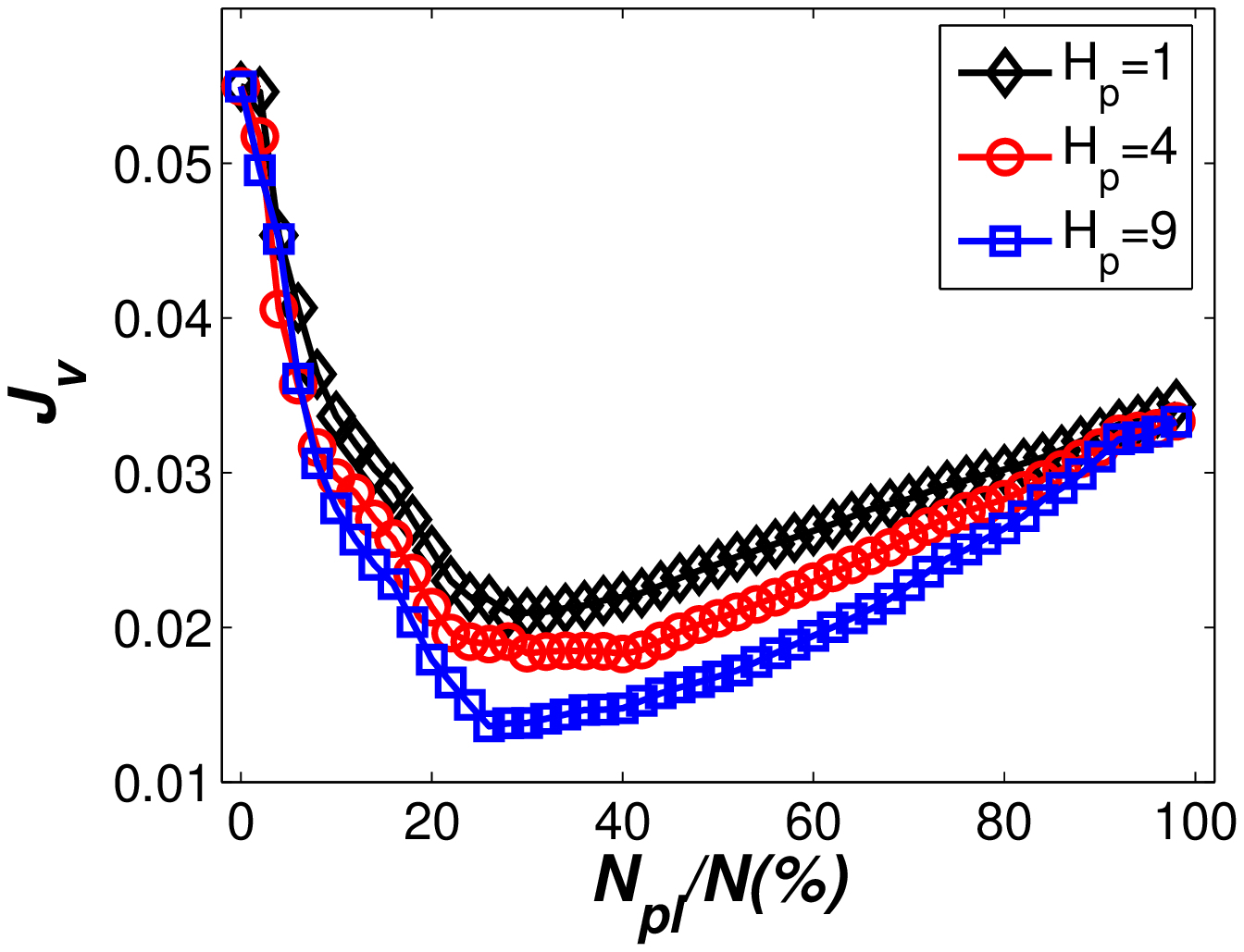}} &
\resizebox{4.3cm}{!}{\includegraphics[width=11.2cm]{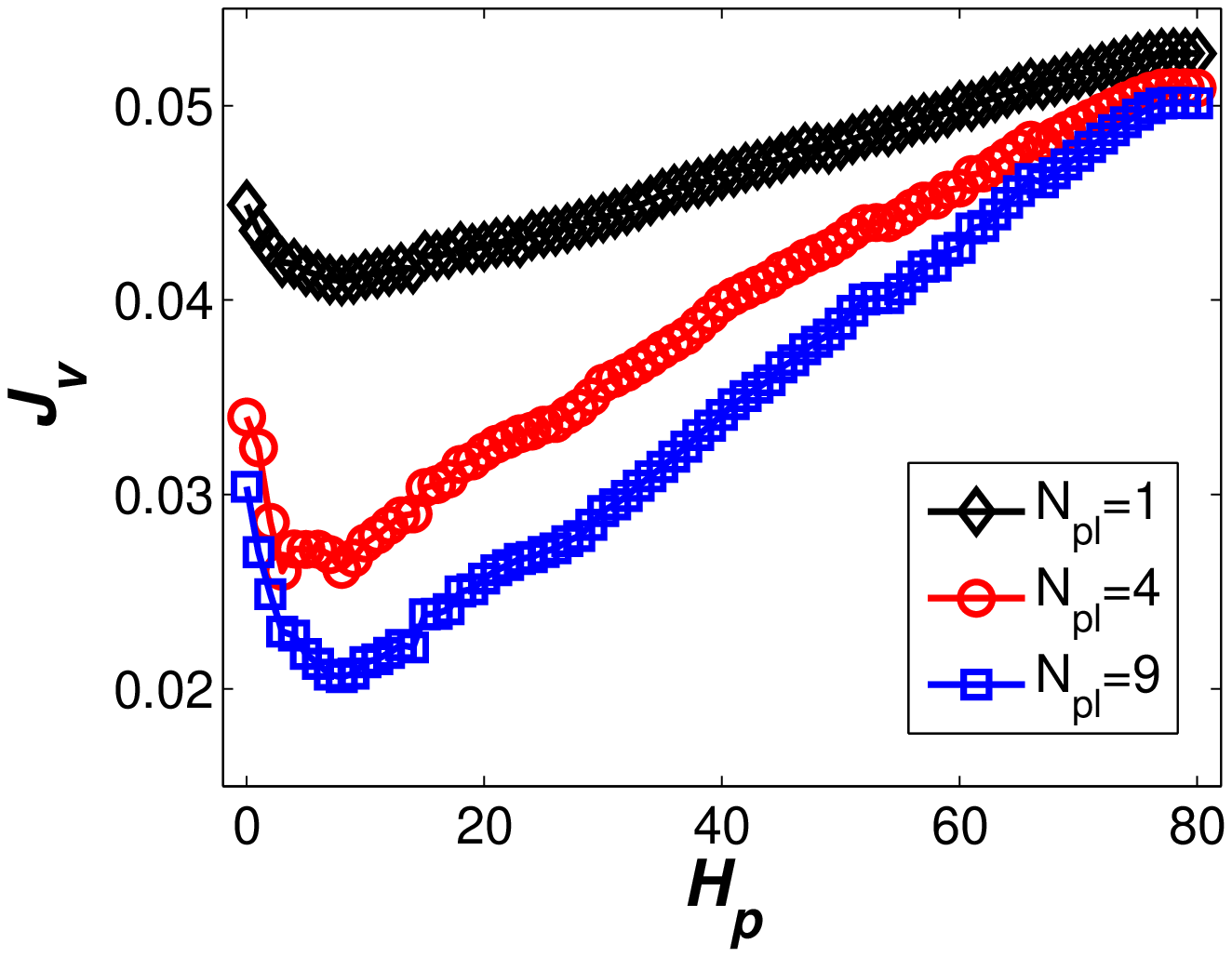}}
 \\
{\scriptsize (a) } &  {\scriptsize (b) } \\
\resizebox{4.3cm}{!}{\includegraphics[width=11.2cm]{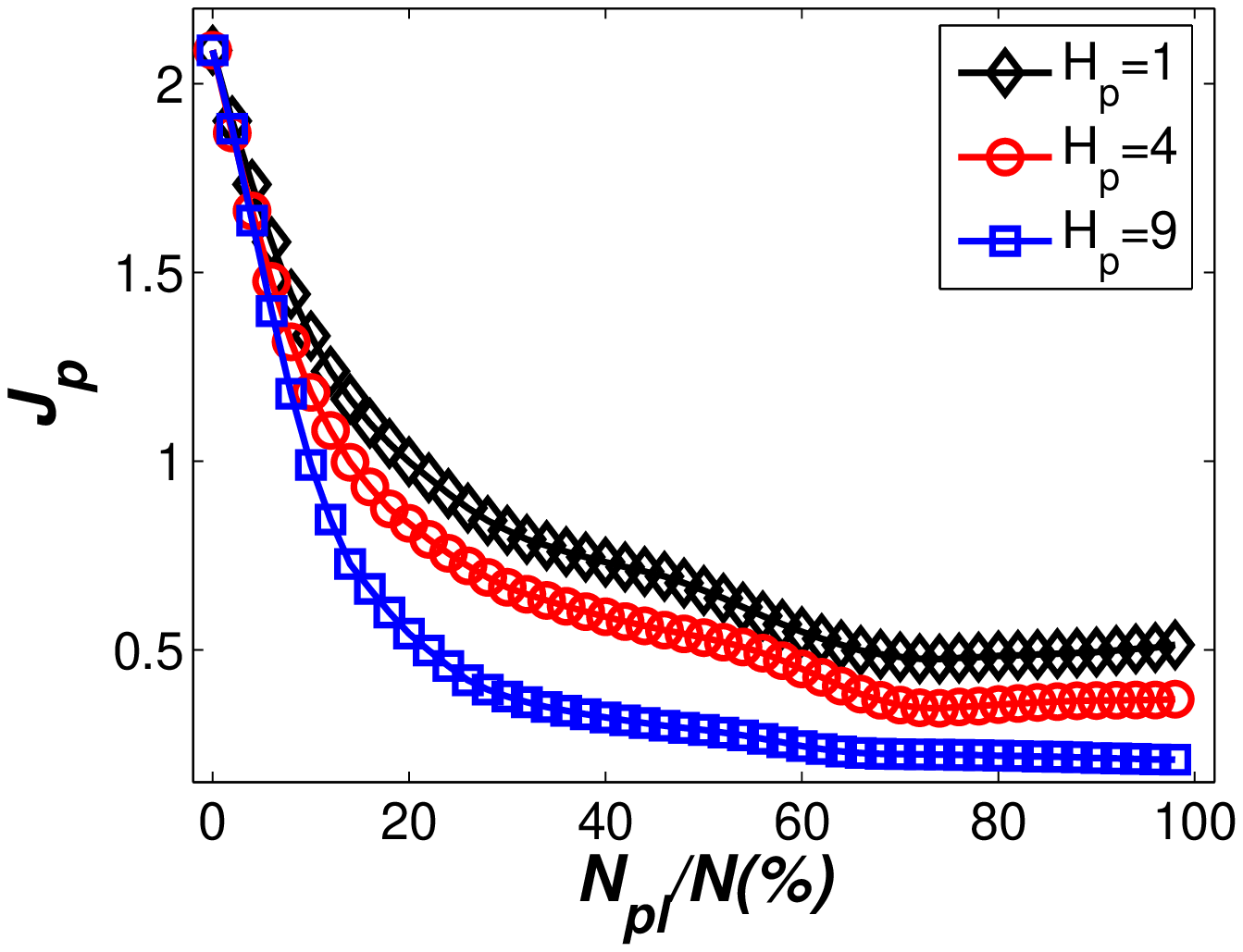}} &
\resizebox{4.3cm}{!}{\includegraphics[width=11.2cm]{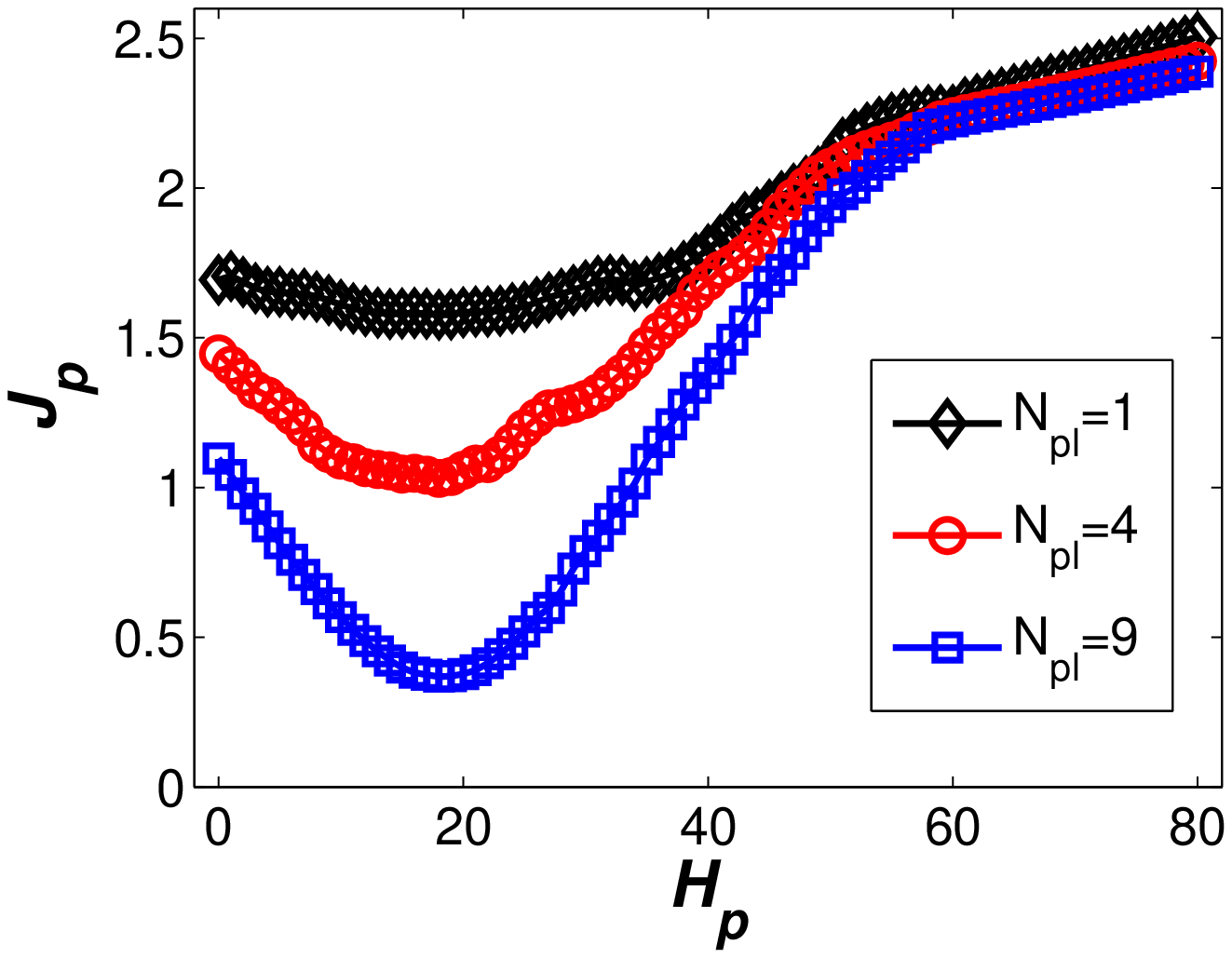}}\\
{\scriptsize (c) } &  {\scriptsize (d)  }
\end{tabular}
\caption{(Color online) The roles of the pseudo-leaders' number
$N_{pl}$ (figures (a) and (c)) and prediction horizon $H_p$ (figures
(b) and (d)) on a flock with a total of $N=50$ agents. The
trajectory of the leader is set along the parabolic curve defined by
$x_2=x_1^2$, and the velocity of the leader is
$v_{L_{x_1}}(t)=0.02,~v_{L_{x_2}}(t)=(0.02(t+1))^2-(0.02t)^2$. The
other parameters and initial conditions are the same as those in
Fig.~\ref{fig:the roles of Hp and pseudo-leaders}}
\end{figure}

The conclusions drawn in this paper are not sensitive to the
trajectory of the leader. In order to validate the generality of the
conclusions on the role of predictive mechanisms, we have used
another two different trajectories, i.e., sinusoidal
($x_2=\sin(2x_1)+1$) and parabolic ($x_2=x_1^2$) curves. As shown in
Fig. 9 and Fig. 10, no matter what the leader's trajectory
is, the influences of $H_p$ and $N_{pl}$ on the principal tendencies
of both $J_p$ and $J_v$ remain the same. Those simulations suggest
that our main conclusion, i.e., predictive capability and long-range
links can compensate for the insufficiency of each other, is also
valid in those two cases.

\section*{Appendix C: Effects of prediction errors}
\begin{figure}[htp]
\centering \leavevmode
\begin{tabular}{cc}
\resizebox{4.3cm}{!}{\includegraphics[width=11.2cm]{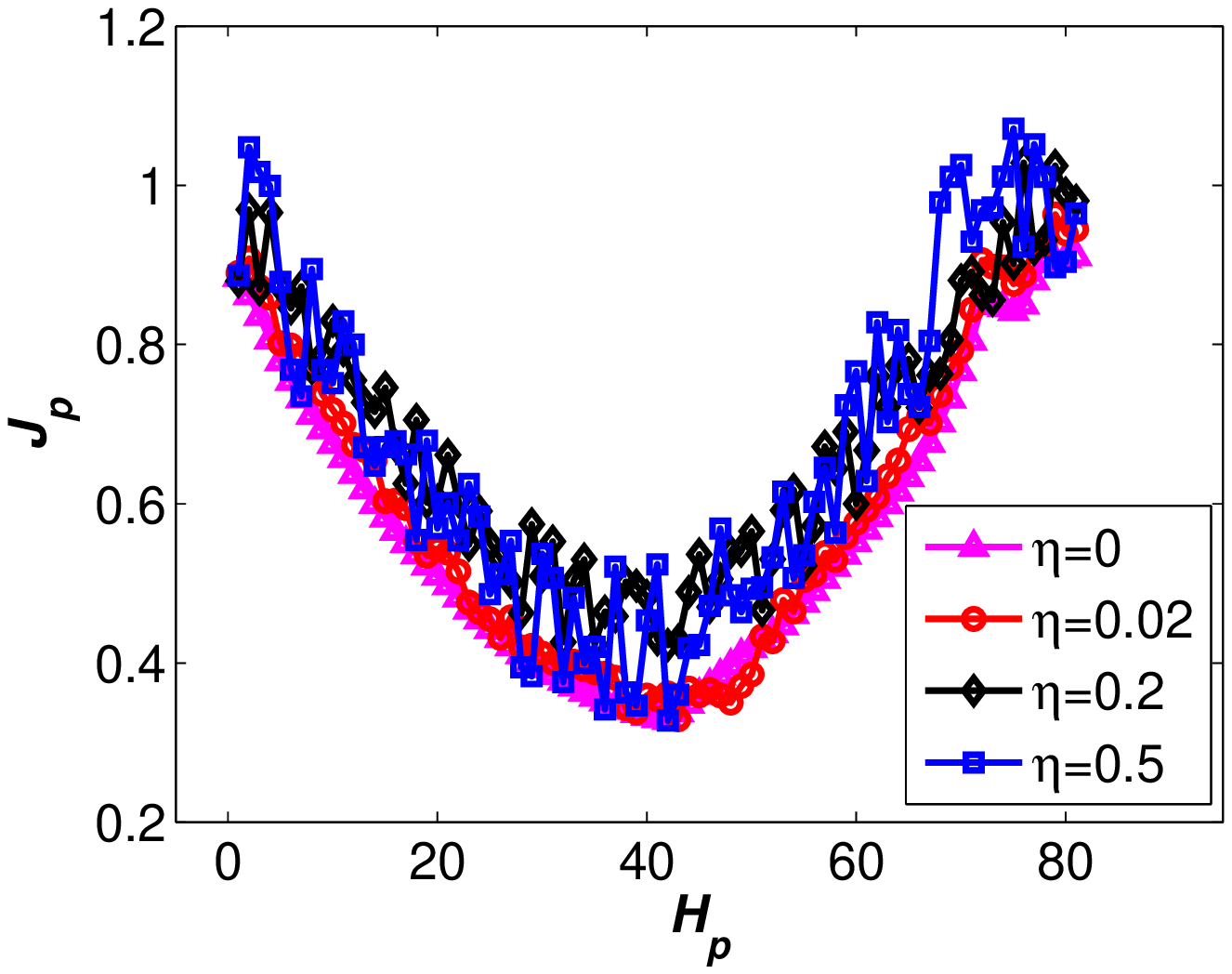}}
&
\resizebox{4.3cm}{!}{\includegraphics[width=11.2cm]{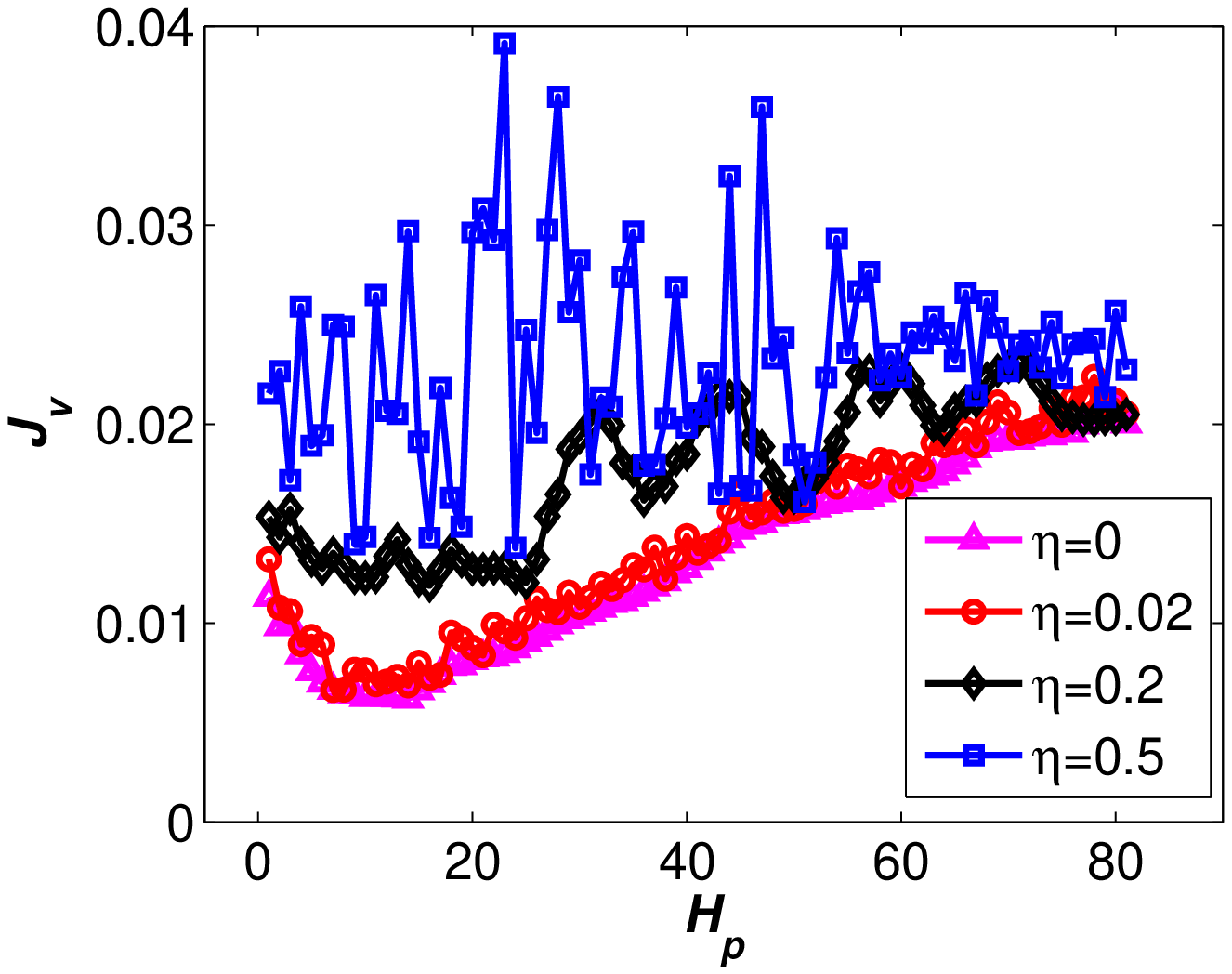}}
 \\
{\scriptsize (a) } &  {\scriptsize (b) } \\
\resizebox{4.3cm}{!}{\includegraphics[width=11.2cm]{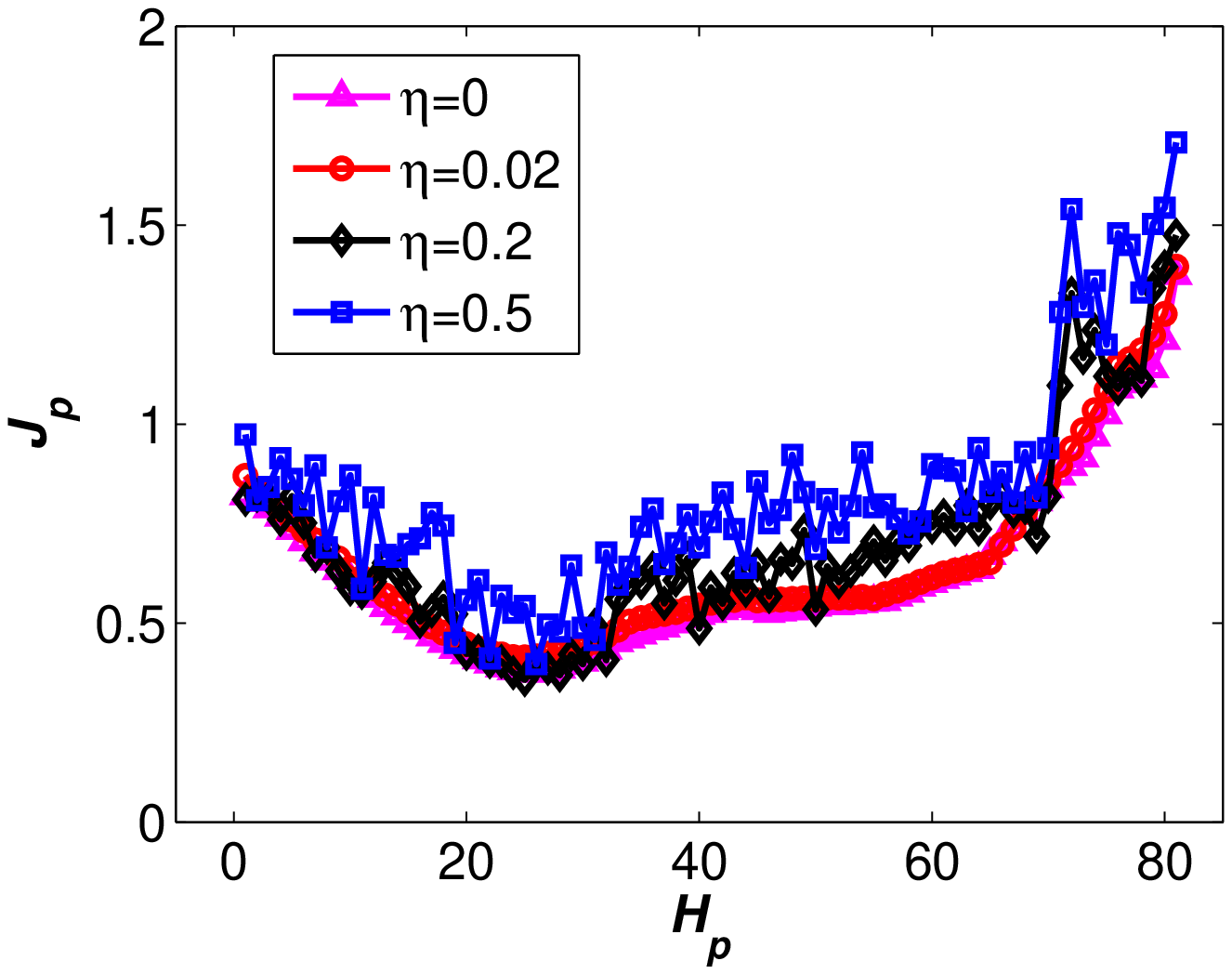}}
&
\resizebox{4.3cm}{!}{\includegraphics[width=11.2cm]{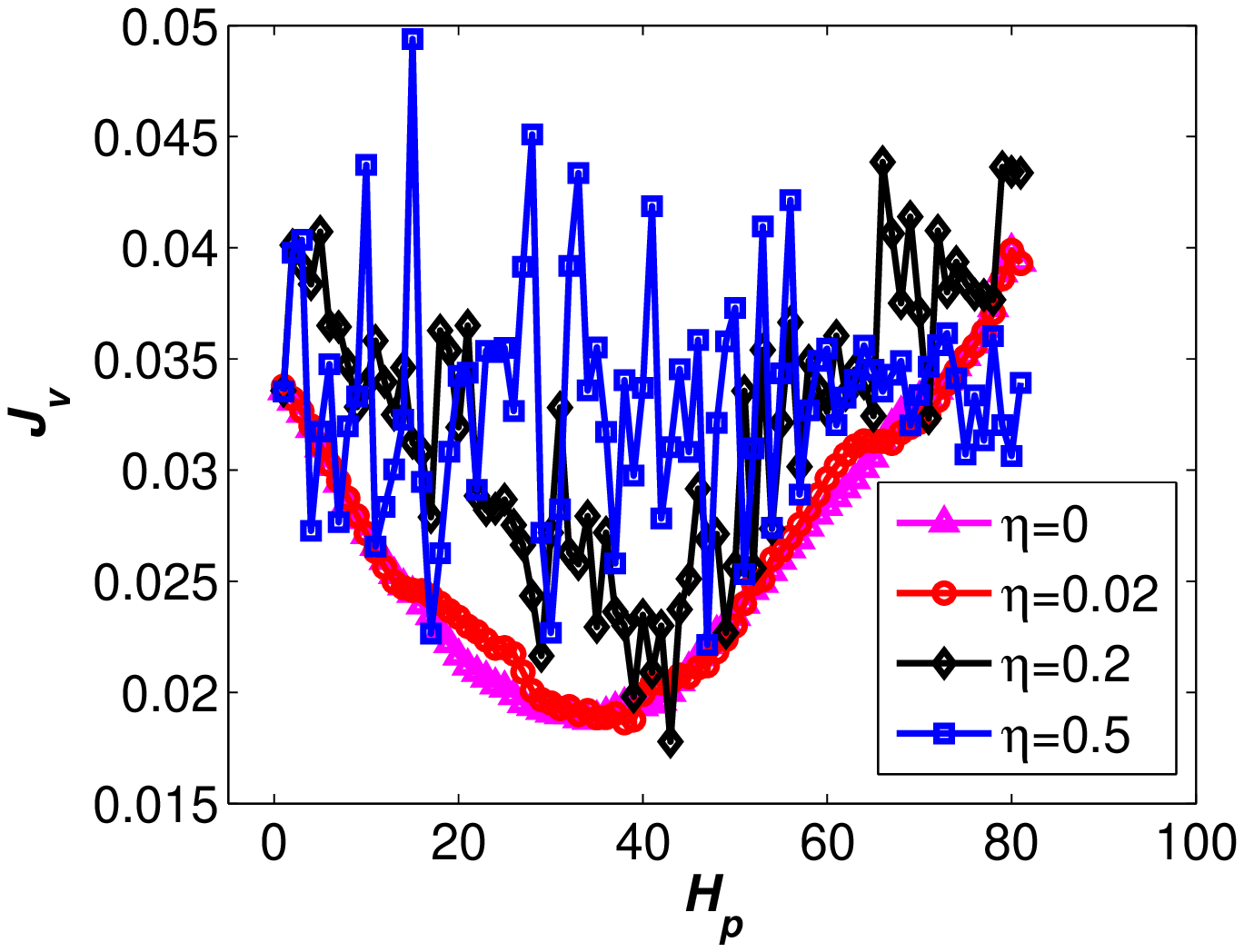}}\\
{\scriptsize (c) } &  {\scriptsize (d) }\\
\resizebox{4.3cm}{!}{\includegraphics[width=11.2cm]{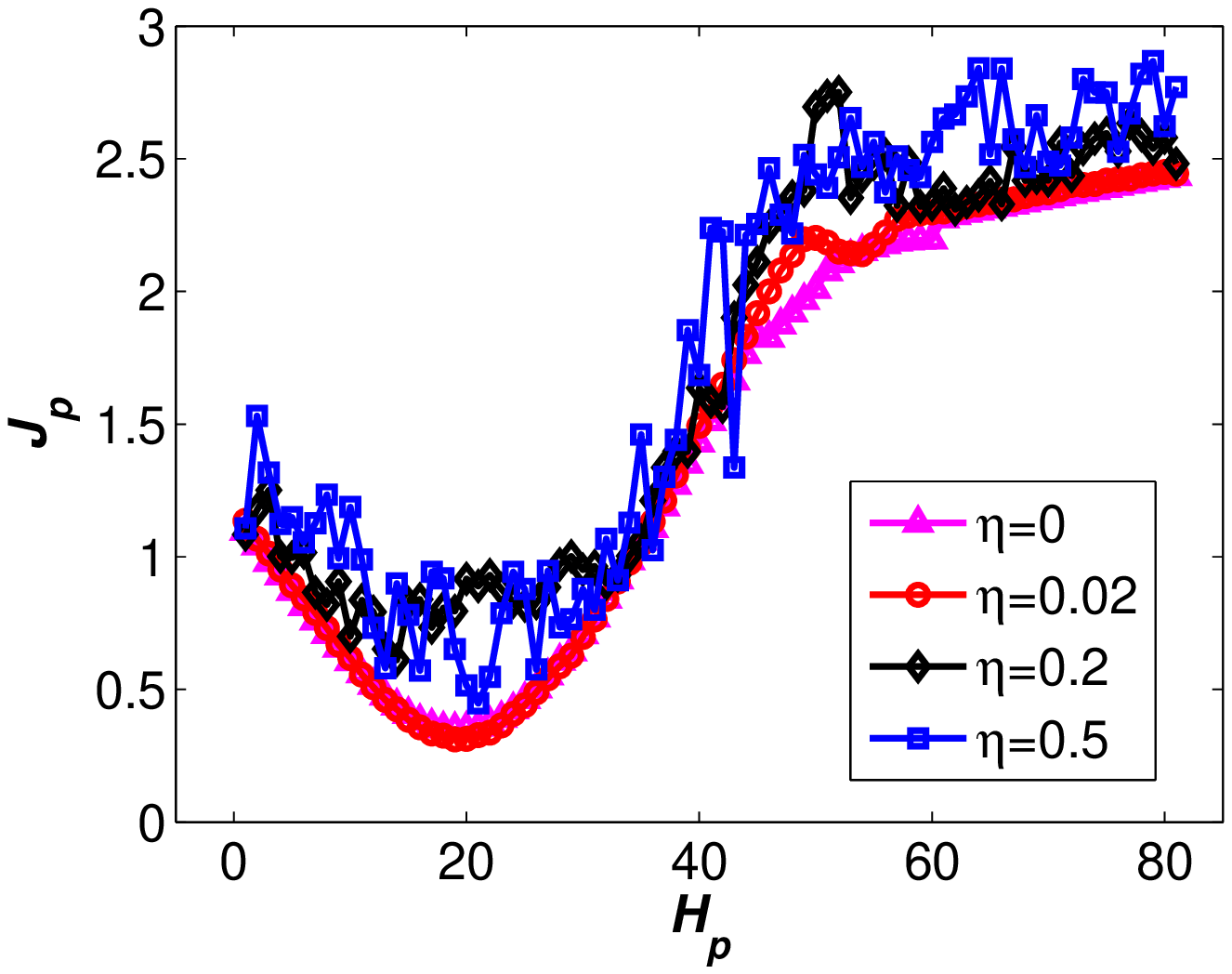}}
&
\resizebox{4.3cm}{!}{\includegraphics[width=11.2cm]{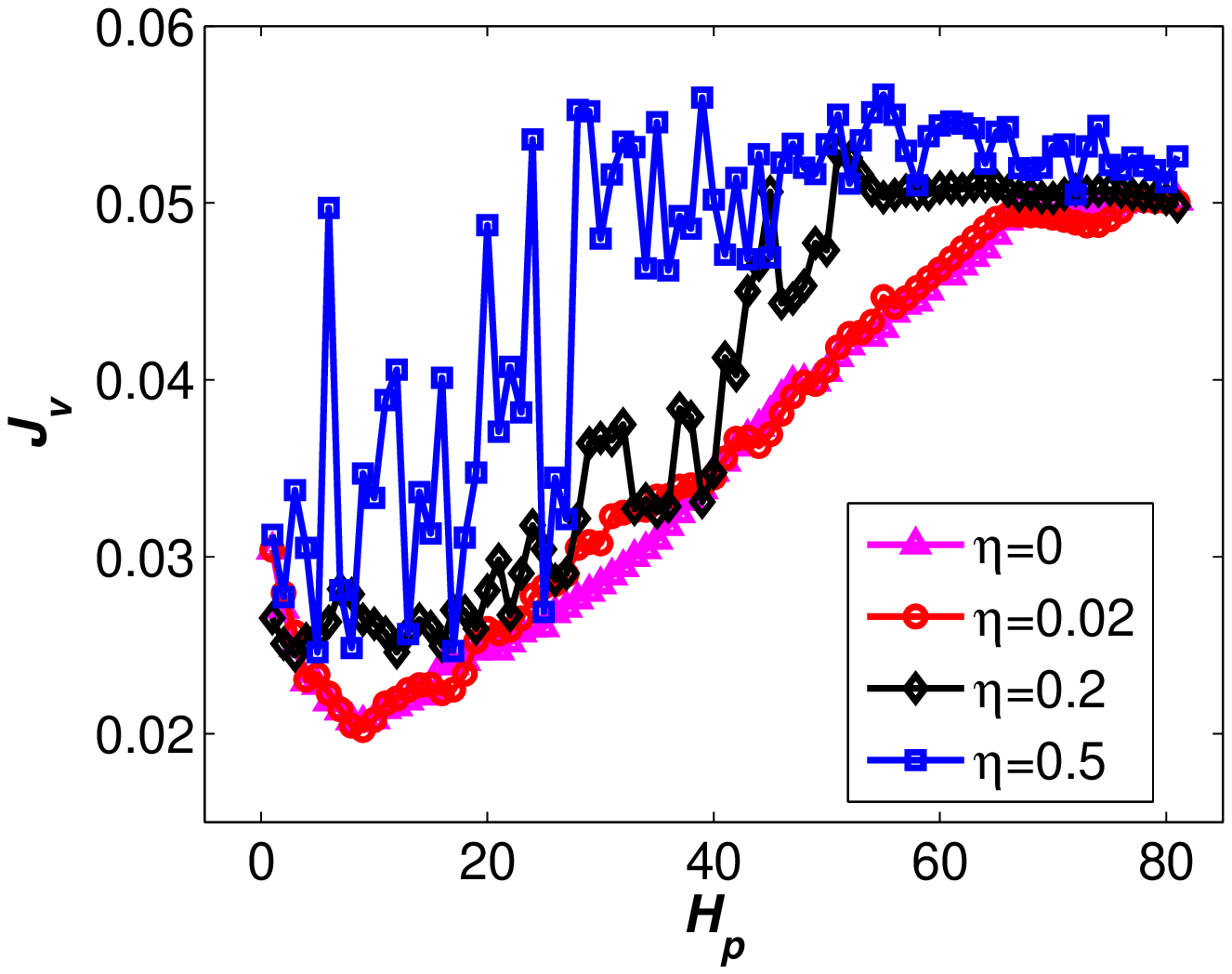}}\\
{\scriptsize (e) } &  {\scriptsize (f) }
\end{tabular}
\caption{(Color online) The effects of prediction error $\xi$. Here,
the trajectory of the leader is set along the curve defined by
$x_2=\sqrt{x_1}$ (sub-figures (a) and (b)), $x_2=\sin(2x_1)+1$
(sub-figures (c) and (d)) and $x_2=x_1^2$ (sub-figures (e) and (f)),
respectively. The flock size is $N=50$, and the number of pseudo-leaders is $N_{pl}=9$.
The other parameters and the initial conditions are the same as
those in Fig.~5. Each point is  averaged over
1000 independent runs. It can be found that moderate prediction
error $\xi$ does not change the global behavior of the flock. If
$\xi$, however, reaches a very large value, like $\eta=0.5$,
the benefits of prediction capability on the synchronization $J_v$
will almost vanish. Note that, the curve of $J_v$ is more sensitive
than that of $J_p$, making the feasible prediction error range of
$J_v$ smaller than the counterpart of $J_p$.}\label{fig:error
effect}
\end{figure}

In realistic system models, mismatch and external perturbations
which cause prediction errors are always present. For completeness, we hereby examine the effects of the
prediction errors on synchronization behavior of the flock. As shown
in Fig.~\ref{fig:error effect}, we present curves of $J_v$ and $J_p$
with increasing noise magnitude $\eta=0,0.02,0.2,0.5$ for the cases
of three different leader trajectories (square root, parabolic and
sinusoidal curves), respectively. It is observed that moderate
prediction error $\xi$ ($\xi\leq 0.2$, for example) can hardly
change the synchronization tendency of the flock, which
guarantees the feasibility and superiority of the current predictive mechanism. On the contrary, too large
$\xi$ ($\xi\geq 0.5$, for example) will inevitably impair the
advantages of this predictive mechanism. This observation is
reasonable, since every method has its own limits and one can not
expect a poor prediction capability to yield a superb guidance for
the flock. Fortunately, the tolerance range of prediction error is
satisfactorily large, which further verifies the generality of our
proposed predictive mechanism. Note that the curve of $J_p$ is more
robust than the curve of $J_v$, which roots in that the relative
velocities can be more easily deviated by prediction error than the
relative positions.


\begin{thebibliography}{99}
\bibitem{ex1} In this paper, a trajectory is considered to be
``regular" if it preserves a continuous, differentiable curves, like
parabolic, sinusoidal and circular waves, along with the increasing
time variable $t$. Moreover, it should not contain any large
fluctuation or some chaos-like movement so that its future tendency
can be predicted by smoothing the historical sequence.  For
instance, when catching a frisbee, a dog will make prediction on its
historical parabolic trajectory, and head for the predicted point of
fall. In this case, the trajectory of the frisbee is ``regular", and
this kind of examples often happen in the real world.
\end{thebibliography}
\end{document}